\documentclass[aps,prd,showpacs,eqsecnum,twocolumn,superscriptaddress]{revtex4}

\usepackage{latexsym} \usepackage{amssymb} \usepackage{amsfonts}
\usepackage{amsmath} \usepackage{bm} \usepackage[dvips]{graphicx}
\usepackage{color}
\usepackage{subfigure} \usepackage{times} \usepackage{units}
\usepackage{hyperref}
\usepackage{bm}
\usepackage[utf8x]{inputenc} \usepackage{amssymb,amsmath}
\usepackage{graphicx} \usepackage[squaren]{SIunits}

\newcommand{\balpha}{\bm{\alpha}} 
\newcommand{\bD}{\textbf{D}}

\begin{document}
\title{Denoising of gravitational wave signals via dictionary learning algorithms}
\author{Alejandro \surname{Torres-Forn\'{e}}}\affiliation{Departamento de
  Astronom\'{\i}a y Astrof\'{\i}sica, Universitat de Val\`encia,
  Dr. Moliner 50, 46100, Burjassot (Val\`encia), Spain} 

\author{Antonio \surname{Marquina}}\affiliation{Departamento de
  Matem\'aticas, Universitat de Val\`encia,
  Dr. Moliner 50, 46100, Burjassot (Val\`encia), Spain}  
  
\author{Jos\'e A. \surname{Font}}\affiliation{Departamento de
  Astronom\'{\i}a y Astrof\'{\i}sica, Universitat de Val\`encia,
  Dr. Moliner 50, 46100, Burjassot (Val\`encia), Spain}
    \affiliation{Observatori Astron\`omic, Universitat de Val\`encia, C/ Catedr\'atico 
  Jos\'e Beltr\'an 2, 46980, Paterna (Val\`encia), Spain}
  
\author{Jos\'e M. \surname{Ib\'a\~nez}}\affiliation{Departamento de
  Astronom\'{\i}a y Astrof\'{\i}sica, Universitat de Val\`encia,
  Dr. Moliner 50, 46100, Burjassot (Val\`encia), Spain} 
    \affiliation{Observatori Astron\`omic, Universitat de Val\`encia, C/ Catedr\'atico 
  Jos\'e Beltr\'an 2, 46980, Paterna (Val\`encia), Spain}


\begin{abstract}
Gravitational wave astronomy has become a reality after the historical detections accomplished during the first 
observing run of the two advanced LIGO detectors. In the following years, the number of detections is expected to
increase significantly with the full commissioning of the advanced LIGO, advanced Virgo and KAGRA detectors.
The development of sophisticated data analysis techniques to improve the opportunities of detection for low 
signal-to-noise-ratio events is hence a most crucial effort. We present in this paper one such technique, 
dictionary-learning algorithms, which have been extensively developed in the last few years and successfully 
applied mostly in the context of image processing. However, to the best of our knowledge, such algorithms have 
not yet been employed to denoise gravitational wave signals. By building dictionaries from numerical relativity 
templates of both, binary black holes mergers and bursts of rotational core collapse, we show how 
machine-learning algorithms based on dictionaries can be also successfully applied for gravitational wave 
denoising. We use a subset of signals from both catalogs, embedded in non-white Gaussian noise, to assess our
techniques with a large sample of tests and to find the best model parameters. The application of our method to the 
actual signal GW150914 shows promising results. 
Dictionary-learning algorithms could be a complementary addition to the gravitational wave data analysis toolkit. 
They may be used to extract signals from noise and to infer physical parameters if the data are in good enough agreement 
with the morphology of the dictionary atoms.
\end{abstract}

\maketitle

\section{Introduction}
\label{section:intro}

The epoch-making detections of the transient gravitational-wave (GW) signals  
GW150914 and GW151226 during the first
observing run of the two Advanced LIGO 
interferometers~\cite{GW150914-prl,GW151226-prl} 
has marked the start of GW astronomy. 
GW150914, detected with unexpectedly high signal-to-noise (SNR) ratio 
(SNR$\sim$24) and with a statistical significance greater than $5.1\sigma$, 
is in excellent agreement with numerical relativity 
waveforms~\cite{SXS, Mroue:2013, campanelli:2006} 
for the final few cycles (chirp), merger (burst) and  
subsequent ringdown of the coalescence of two stellar-origin 
black holes (BHs) in a binary system.  
GW151226, also the result of a binary BH merger, 
was recovered with similar statistical significance but with a SNR$\sim$13. 
Its initial BH masses, $14.2M_\odot$ and $7.5M_\odot$, 
are lower than in the case of GW150914, $35M_\odot$ and $30M_\odot$. 
As a result, GW151226 spent almost 1 s in the LIGO frequency band, 
increasing in frequency and amplitude from 35 to 450 Hz over about 55 cycles. 
Contrary to GW150914, matched filtering with waveform templates 
from general relativity was essential to detect GW151226 
due to the smaller strain amplitude and the longer time 
interval~\cite{GW151226-prl}.

At present, the two Advanced LIGO interferometers are being upgraded. 
The second observing run (O2) is expected to start in late 2016 
with a significant strain improvement. 
At the same time, the commissioning of the European detector 
Advanced Virgo~\cite{Virgo} is well underway, aiming at start 
observing in the second half of this year, while the Japanese detector 
KAGRA~\cite{KAGRA} is still under construction. 
Simultaneous observational campaigns of these four detectors, 
five with the later addition of the recently approved LIGO India, 
will increase considerably the rate of detections along with 
their statistical significance and the accuracy of the sky location 
of each event~\cite{abbott-lrr}.

Despite the recent discoveries, noise removal remains one of the 
most challenging problems in GW data analysis.
There exist a number of noise sources that limit the 
possibilities of detection~\cite{Martynov:2016}.
The most limiting source of noise for frequencies below a few tens 
of Hz is { gravity gradient} noise. Thermal noise due to
Brownian motion is dominant at intermediate frequencies, 
while shot noise, produced by quantum fluctuations of 
the laser, becomes prominent at frequencies above 
$\sim 150$ Hz, { difficulting} detection above 2 kHz. 
{ Nevertheless, searches for gravitational wave bursts up to frequencies of 5 kHz have been 
performed~\cite{Abadie:2012}.}
These sources of noise are not stationary and the sensitivity 
of the detectors changes with time. 
To add more complexity, transient spurious noise signals (glitches) 
due to instrumental or environmental sources, may potentially 
disturb astrophysical signals. 
Glitches might mimic GW signals increasing the false alarm rate 
and producing a decrease in the detectors' duty cycles.
A huge effort in commissioning and detector 
characterization~\cite{Christensen:2010} 
has been done to reduce the effect of glitches. 
{ Improving glitch identification and classification~\cite{Powell:2015,Powell:2016} would improve 
detection efficiency but there will always be a chance for false positives 
in the detectors.}

GW detectors are designed to be sensitive to waveforms produced by different 
astrophysical mechanisms. Sources can be separated in groups depending on 
how well-known and modeled their waveforms are. 
Specific data analysis techniques have been developed for each type of signal 
(for a review see~\cite{Jaranowski:2012} and references therein). 
Transient GW signals from compact binary coalescence (CBC), 
either from binary neutron stars (BNS) or binary black holes (BBH), 
are well studied and the corresponding waveforms 
can be calculated with high accuracy. 
These systems are typically modeled using the effective-one-body formalism (EOB) 
\cite{Buonanno:2000} which combines post-Newtonian methods \cite{Blanchet:1995, Blanchet:2004} 
with numerical relativity and perturbation theory \cite{Nagar:2005}. 
This technique allows to generate template banks efficiently.
This is the main reason matched-filtering is the most common method for CBC 
detection~\cite{Owen:1999, Schutz:2009,1606.04856} in which filters correlate signals with templates. 
A trigger associated with a specific template is generated when the filter output 
excess a certain threshold. 
{ In addition to EOB waveforms, other waveform families are needed to cover as much
parameter space as possible (see~\cite{prl241102} and references therein).}

Matched-filtering becomes however unpractical for well-modeled but continuous sources, 
like spinning neutron stars, due to the large computational resources it would require. 
Nevertheless, as such signals are very stable and have long duration, 
a coherent integration can be performed. 
In addition, the data from all detectors can be compared, 
which increases the SNR of this type of events. 
Roughly speaking there are two main methods to fulfill this comparison, 
cross-correlation methods and coherent methods~\cite{bose:2000,klimenko:2008}. 
The former  directly  compare  the  data streams from a pair of detectors 
to search for a common signal within  uncorrelated  noise while the latter 
generalize the concepts of excess power and cross-correlation to take full 
advantage of having three or more data streams. 
The duration and the sky coverage (all-sky or targeted)~\cite{palomba:2012} 
can vary depending of the type of source which is sought for.

In contrast with the last type of sources, 
the non-spherical gravitational collapse of massive stars 
produces a short ($\sim$ms) duration (prompt) signal 
(but see~\cite{cerda-duran15} for the case of collapsars where the duration of 
the signal is dominated by the accretion timescale, considerably longer) with a significant power 
in the kHz frequency band. { In addition to core-collapse supernova, other 
astrophysical sources as cosmic string cusps~\cite{damour:2005} and BBH mergers, 
can also produce GW transients or ``bursts".}
{ Such signals, in particular core-collapse bursts,}  can only be modeled imperfectly, 
and the computational requirements for obtaining the corresponding waveforms 
from numerical relativity simulations are much larger than in the case of CBC. 
Therefore, a bank of templates cannot be built with sufficient accuracy to meet 
the  requirements of matched-filtering. For burst signals, the time-frequency analysis 
of the signal in all the detectors,  related to each other with cross-correlation 
and coherent methods, is the best option. { With initial detectors, 
a complete all-sky, all-time burst search was performed~\cite{Abadie:2012} and
has also been carried out during Advanced LIGO's first observing run (O1)~\cite{Abbott:20162}.}
These techniques, used in tandem with electromagnetic observations, 
can increase the possibilities of identifying a GW burst \cite{Abbott:2009}. 

{
The detection confidence of unmodeled astrophysical sources has
significantly improved in recent years. In particular, coherent approaches 
over a network of GW detectors have proven to be very 
effective~\cite{Thrane:2015,Klimenko:2016}, increasing the detection 
confidence of long-duration (above several seconds) burst signals which are 
insensitive to the 
presence of most noise transients. In contrast, short-duration bursts are more 
affected by detector glitches and specific pipelines based on Bayesian 
inference have been developed to differentiate between signals and 
noise transients, namely {\tt coherentWaveBurst}~\cite{Kanner:2016},
 {\tt BayesWave}~\cite{Littenberg:2016}, and {\tt oLIB}~\cite{Lynch:2016}.
Other approaches, like those of~\cite{Rover:2009,Engels:2014},
have proven to be effective for estimating physical parameters and for
the reconstruction of burst signal waveforms from (Gaussian) noisy
environments.
}

Apart from the standard techniques mentioned in the previous paragraphs, 
GW data analysis can benefit from the incorporation of new approaches from
other fields. Recently, we presented in~\cite{Torres:2014} new methods for denoising GW signals 
based on $L^1$-norm minimization, modeling the denoising problem as a variational problem. 
These methods have been originally developed and fully tested in the context of 
image processing where they have been shown to be the best approach to solve the 
so-called Rudin-Osher-Fatemi denoising model~\cite{Rudin:1992}. 

In this paper we continue the work initiated in~\cite{Torres:2014}.
We assume the linear degradation model
to solve the denoising problem as
the estimation of the recovered signal $\textbf u$ from the relation
$\textbf f=\textbf u+{\textbf n}$ where the measured signal is
$\textbf f$ and $\textbf n$
is white Gaussian noise of zero mean.
Our approach in~\cite{Torres:2014} consisted in obtaining
$\textbf u$ as the unique minimizer
of the total variation norm of the signal subject to a fidelity term expressed
in terms of the $L^2$-norm of the residual, i.e.,
$||\textbf f- \textbf u||_{L^2}^2$.
Here, we propose an alternative approach based on the sparse
reconstruction of the signal $\textbf u$ over learned dictionaries,
built from existing waveform catalogs.
The development of models and algorithms for sparse reconstruction of signals
over a dictionary has been a subject of great interest in recent 
years~\cite{Chen:2001,Elad:2006,Mairal:2013}.
It appears as an alternative to the traditional signal representation
based on Fourier decomposition or more
modern representations based on wavelets,
chirplets, warplets, etc.

Following the terminology introduced by
Mallat and Zhang~\cite{Mallat:1993}
a {\it dictionary} is a collection of signals
(in our case waveforms) 
of length $n$ called {\it atoms}, 
\begin{equation}
\label{eq:dict_decomp}
{\textbf u}= \bD \bm{ \alpha},
\end{equation}
where $\textbf u$ is the signal to be recovered, ${\bD}=[\bf{d_1}, \ldots ,\bf{d_p}]$ 
is the dictionary, which is composed of $p$ atoms of length $n$, 
and $\balpha\in\mathbb{R}^p$ is a vector which contains the coefficients of the representation. 
The atoms are not set to be orthogonal unlike in other decompositions like 
those based in Principal Component Analysis (PCA), allowing more flexibility in the representation. 
A dictionary can be complete, if it contains exactly $p=n$ atoms, 
or overcomplete if $p>n$ atoms. In the latter case
the solution vector $\balpha$ is not unique and cannot be obtained 
by applying simple linear methods. 
In our work, we use the {\it basis pursuit decomposition} proposed in~\cite{Chen:2001}. 
An interesting review of other approaches to solve 
problem (\ref{eq:dict_decomp}) can also be found in Ref.~\cite{Chen:2001}.

The prototype signals of a dictionary can be chosen as a predefined set of functions, 
like a Fourier basis (frequency dictionaries), 
several types of wavelet functions (wavelet dictionaries) 
or Gabor wavelet decomposition to produce time-frequency dictionaries.
However, the idea of using a dictionary learned from data has improved 
the denoising results considerably~\cite{Elad:2006}.
Nowadays, state-of-the-art algorithms for reconstruction and denoising are 
being developed along this direction~\cite{Mairal:2013}, and 
very efficient methods have been devised to solve the challenging 
optimization problem inherent to learning dictionaries. 
  
The sparse reconstructions of signals over 
trained dictionaries we propose in this work are obtained for the same kind of 
GW trained signals that we did in~\cite{Torres:2014}, namely burst signals from a catalog of rotational stellar 
core collapse~\cite{Dimmelmeier:2008} and chirp-burst-ringdown signals 
from a catalog of BBH mergers~\cite{Mroue:2013}. 
Once the dictionaries are set, we demonstrate their utility for the denoising 
of GW signals embedded in Gaussian noise. 

This paper is organized as follows: Section~\ref{section:math} 
describes the basic mathematical details of our method, namely
the sparse representation and the dictionary learning problems 
together with the specific formulation we use to solve them. 
Section~\ref{section:gw} deals with the GW waveform catalogs 
we employ to assess our method. 
In Section~\ref{section:parameter_estimation} we adapt the 
general problem to the specific case of GW signals and we obtain 
the optimal set of model parameters to perform the denoising of given signals. 
In Sections~\ref{section:application} and~\ref{section:other} we illustrate our technique 
with a significant sample of test cases. 
Section~\ref{section:GW150914} discusses the performance of our method 
when applied to the actual signal GW150914. 
Finally, the conclusions of our work and possible future extensions are 
presented in Section~\ref{section:summary}.  
Appendix A contains a table with the correspondence between the naming of the 
GW signals employed in this study and that of the original GW catalogs.

\section{Mathematical framework}
\label{section:math}
\subsection{Sparse Reconstruction over a fixed dictionary}

We start from the linear degradation model
\begin{equation}
{\textbf f}={\textbf u}+{\textbf n},
\end{equation}
where $\bf{f}$ is the measured signal, $\textbf{u}$ is the GW signal to be recovered 
and $\textbf{n}$ is random Gaussian noise. 
If signal $\textbf{u}$ is a vector $\textbf{u}\in\mathbb{R}^n$ one can say that 
it admits a sparse approximation over an overcomplete dictionary 
$\textbf{D}\in \mathbb{R}^{n\times p}$, where each column contains one of the $p$ atoms 
of length $n$ and $p>n$, when one can find a linear combination of a {\it few} atoms 
from $\bD$ that is close to $\textbf{u}$. 
The classical dictionary learning techniques \cite{Olshausen:1997, Aharon:2006} 
try to solve the variational problem associated with Eq.~(\ref{eq:dict_decomp}) 
given by,
\begin{equation}
 \balpha=\underset{\balpha}{\rm{argmin}} ||\balpha||_0\,\,~\rm{subject~to}\,\,~\bD\balpha\sim {\textbf u},
\end{equation}
where $||\cdot||_0$ is the $L^0$-norm to assure that we have the solution 
with the fewest number of nonzero coefficients. The $L^0$-norm is just the number
of nonzero components of the vector.
This constrained variational problem can be formulated as an unconstrained 
variational problem adding the  $L^0$-norm term as a penalty term weighted 
by a Lagrangian multiplier $\lambda$,
\begin{equation}
\label{eq:l0-problem}
\balpha=\underset{\balpha}{\rm{argmin}} ||\bD\balpha- {\textbf f}||^2_2+\lambda||\balpha||_0.
\end{equation}
This problem is not convex and is NP-hard (i.e.~non-deterministic polynomial-time hard) 
so, in practice, it cannot be solved in linear time~\cite{Marcellin:2000}. 
A problem is in the NP class if it can be solved in non-deterministic polynomial-time. 
Algorithms that produce and approximate solutions to this problem have been 
proposed in the past. 
The simplest ones are matching pursuit (MP) and 
orthogonal matching pursuit (OMP) (see~\cite{Chen:2001} 
and references therein for details).

The variational problem defined by Eq.~(\ref{eq:l0-problem}) can be reformulated 
into a convex variational formulation 
by substituting the $L^0$-norm by the nondifferentiable convex $L^1$-norm 
in the total energy. 
The regularization in the $L^1$-norm promotes zeros in the components of the 
vector coefficient $\balpha$. 
This problem can be solved in linear time and the solution found is the 
sparsest one in most cases. The variational problem thus stands as,
\begin{equation}
\label{eq:lasso}
\balpha=\underset{\balpha}{\rm{argmin}} ||\bD\balpha- {\textbf f}||^2_2+\lambda||\balpha||_1,
\end{equation}
which is known as {\it basis pursuit}~\cite{Chen:2001} or {\it LASSO}~\cite{lasso}. 
An alternative formulation known as {\it elastic-net}~\cite{Zou:2005},
\begin{equation}
\label{eq:elastic_net}
\balpha=\underset{\balpha}{\rm{argmin}} ||\bD\balpha- {\textbf f}||^2_2+\lambda_1||\balpha||_1+\frac{\lambda_2}{2}||\balpha||_2^2,
\end{equation}
adds a $L^2$-norm penalty for stability reasons, i.e., the calculated
approximate representation depends on the data as a Lipschitz function.

To solve the LASSO problem (\ref{eq:lasso}) we propose a decomposition based 
on the Split-Bregman (SB) algorithm \cite{Goldstein:2009}. 
The SB algoritm solves very efficiently $L^1$-norm minimization problems, 
like the Rudin-Osher-Fatemi variational model \cite{Rudin:1992}, 
decoupling the $L^1$-norm and $L^2$-norm terms and 
solving them alternatively until convergence is reached. 
In order to achieve this goal we introduce a new independent unknown vector ${\textbf d}$
to split the minimization with respect to the $L^1$-norm.
Applying this splitting the problem reads as follows
\begin{equation}
\label{eq:sb-algorithm}
(\balpha,{\textbf d})=\underset{\balpha, {\textbf d}}{\rm{argmin}} ||\bD\balpha- {\textbf f}||^2_2+\lambda|| {\textbf d}||_1+{\mu}||\balpha-{\textbf d}||_2^2\,.
\end{equation}
By iteratively minimizing with respect to $\balpha$ and 
$ {\textbf d}$ separately,
the SB iterative procedure reads as follows
\begin{eqnarray}
\label{split_bregman_iteration1}
 \balpha^{k+1} &=&\underset {\balpha} {\text{argmin}} \frac{\lambda}{2}|| {\textbf f}-\bD\balpha||_{2}^2
  + \frac{\mu}{2} || {\textbf b}^k + \balpha - {\textbf d}^k||_{2}^2\,\\
\label{split_bregman_iteration2} 
  {\textbf d}^{k+1} &=&\underset { {\textbf d}} {\text{argmin}}~| {\textbf d}|+\frac{\mu}{2}|| {\textbf b}^k + \balpha^{k+1} -  {\textbf d}||_2^2 \,,\\ 
\label{split_bregman_iteration3}
  {\textbf b}^{k+1}&=& {\textbf b}^k+\balpha^{k+1} -  {\textbf d}^{k+1}\,,
\end{eqnarray}
starting with ${\textbf b}^{0}=0$. The role of the auxiliar vector 
${\textbf b}$ is to enforce the unknowns $\textbf d$ and $\balpha$ be
equal when convergence is reached. The iteration process uses a
small positive fixed value of $\mu$  and it runs over a scale-space that
reconstructs the signal as a linear combination of few elements of
the dictionary.
Since the two parts are decoupled, they can be solved independently. 
The energy of the first step is smooth (i.e.~differentiable) and it 
can be solved using common techniques as the Gauss-Seidel method. 
On the other hand, ${\textbf d}$ can be explicitely computed to optimal values 
by using the shrinkage operator,
\begin{equation}
\label{eq:shrink}
 {\textbf d}^{k+1}={\text{shrink}}( {\textbf b}^k+ \balpha^{k+1},1/\mu)~,
\end{equation}
\begin{equation}
\label{eq:shrink2}
{\text{shrink}}(x,\gamma)=\frac{x}{|x|}\text{max}(|x|-\gamma,0)~.
\end{equation}
In practice we only use one iteration for the splitting steps 
and the final algorithm only consists
of just one loop (see \cite{Goldstein:2009} 
for a detailed discussion). 

\subsection{Dictionary Learning Problem}

Up to this point we have considered that the dictionary $\bD$ is fixed 
and we only have to solve the problem of representation. 
As we want to design the dictionary to fit a given set of GW signals, 
we start by considering a finite number of training signals, 
which can be split in $m$ patches of length $n$, 
i.e.~$\textbf{U} = [ {\textbf u}_1,\ldots,  {\textbf u}_m]$ in 
$\mathbb{R}^{n\times m} $. In most common problems, 
the number of training patches $m$ is large compared with the length of each patch, 
$n\ll m$. In general, the number of atoms in the dictionary 
is lower than the number of patches, $p\ll m$, because each signal 
only uses a few elements in $\bD$ for the representation.

To obtain the trained dictionary, we need to add the dictionary matrix $\bD$ 
as a variable in the minimization 
problem
\begin{equation}
\label{eq:dict_learning}
\balpha=\underset{\balpha, \bD}{\rm{argmin}} \frac{1}{n}\sum_{i=1}^{m}||\bD\balpha_i- {\textbf u}_i||^2_2+\lambda||\balpha_i||_1,
\end{equation}
where the summation index $i$ indicates 
the i-th row of $\balpha\in \mathbb{R}^{p\times n}$ (now a matrix), 
which contains the coefficients of the sparse 
representation of each atom in the dictionary. 
The constraint in $\bD$ reads
\begin{equation}
\label{eq:D_constrain}
\bD\in \mathbb{R}^{n\times p}\,\, \rm{subject~to}\,\,\, ({\textbf d}_i^T {\textbf d}_i )\le 1\,\,\,\forall i=1,\ldots, p~.
\end{equation}
The whole problem (\ref{eq:dict_learning}) is not jointly convex, but convex
with respect to either of the two variables, $\balpha, \bD$, keeping the other one fixed.
To perform the dictionary update we follow the algorithm  proposed 
by Mairal et al in~\cite{Mairal:2009} to which
the reader is addressed for details. These authors use a block-coordinate descent method 
\cite{Tseng:2001} for solving for $\bD$ and $\balpha_i$ iteratively,
\begin{eqnarray}
\balpha^{k+1}&=&\underset{\balpha}{\rm{argmin}} \frac{1}{n}\sum_{i=1}^{m}||\bD^{k}\balpha_i- {\textbf u}_i||^2_2+\lambda||\balpha_i||_1\\
\bD^{k+1}&=&\underset{\bD}{\rm{argmin}} \frac{1}{n}\sum_{i=1}^{m}||\bD\balpha_i^{k+1}- {\textbf u}_i||^2_2+\lambda||\balpha_i||_1
\end{eqnarray}
The main advantage of this implementation is that it is parameter-free 
and does not require any learning rate.

\section{Gravitational wave catalogs}
\label{section:gw}

In the present work we employ the same two catalogs of GW signals used in~\cite{Torres:2014} to assess our method, namely a catalog of signals from relativistic rotational core collapse simulations~\cite{Dimmelmeier:2008} and a catalog from BBH simulations~\cite{Mroue:2013}. In addition, we also consider two extra signals, one from a core collapse catalog developed by Abdikamalov et al.~\cite{abdikamalov14} and a BBH signal from~\cite{baker:2007}. These last two signals allow us to investigate the ability of our approach to extract independent waveforms using dictionaries built from atoms that do not contain explicit information on the signals to be denoised.
 \begin{figure*}
  \centering
    \includegraphics[width=89mm]{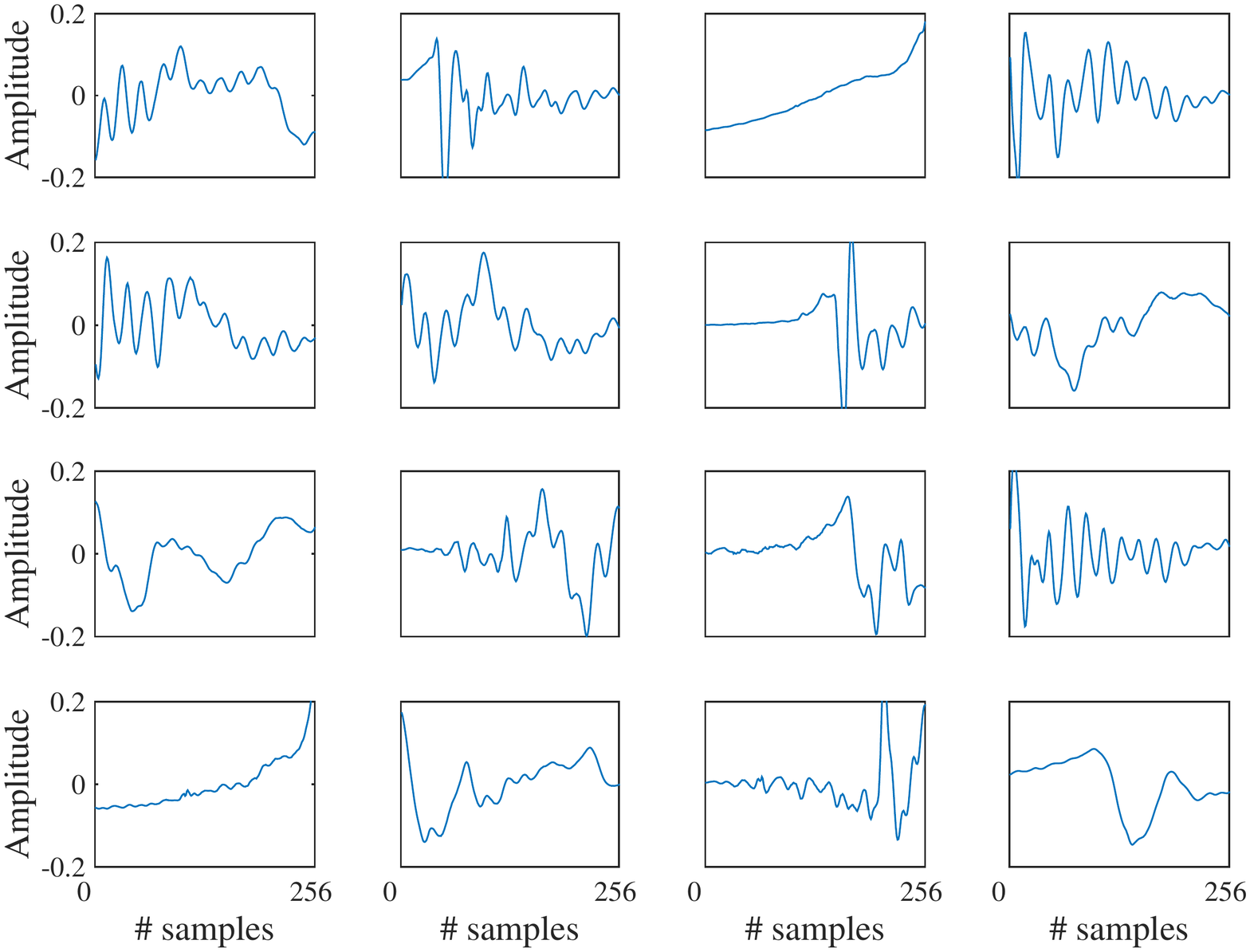}
     \includegraphics[width=89mm]{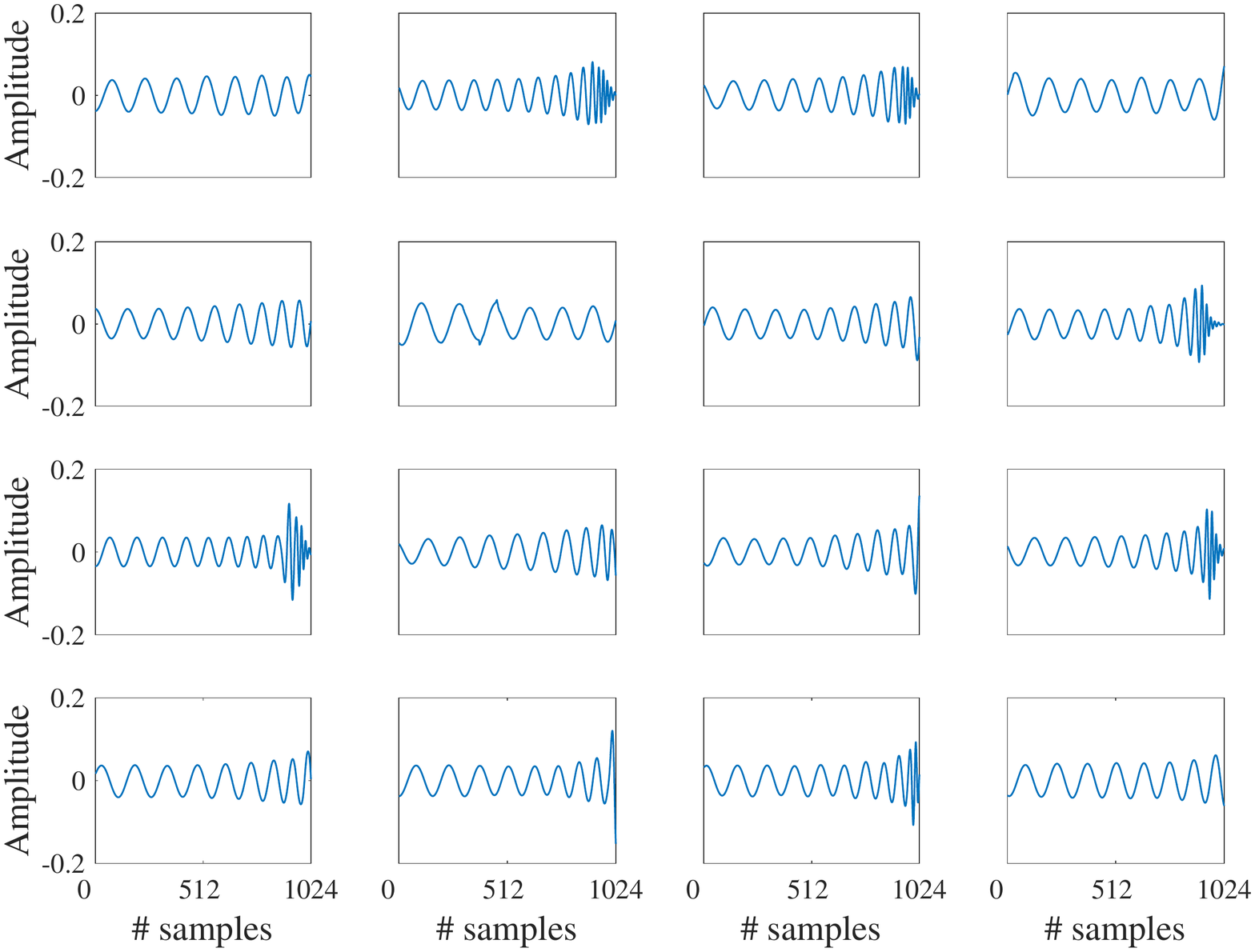}
  \caption{Random examples of the atoms of both dictionaries, for burst waveforms (left panel) and for BBH signals (right panel). The number of samples is shown in the horizontal axis while the normalized amplitude is shown in the vertical axis.}
  \label{fig:dictionary}
\end{figure*}

In the core collapse scenario the bulk of gravitational radiation is emitted during bounce, when the quadrupole moment changes rapidly, which produces a burst of GWs with a duration of about 10 ms and a maximum dimensionless amplitude of about $10^{-21}$ at a distance of $10$ kpc. Broadly speaking, GW signals from core collapse exhibit a distinctive morphology characterized by a steep rise in amplitude to positive values before bounce followed by a negative peak at bounce and a series of damped oscillations associated with the oscillations of the newly formed proto-neutron star around its equilibrium solution. We employ the catalog developed by Dimmelmeier et al.~\cite{Dimmelmeier:2008}, who obtained 128 waveforms from general relativistic simulations of rotating stellar core collapse to a neutron star using the {\tt CoCoNuT} code. The core collapse template bank computed by~\cite{abdikamalov14} has also been built through axisymmetric simulations with the {\tt CoCoNuT} code. The progenitors investigated have different initial angular momentum distributions in the core and the simulations include a microphysical finite-temperature equation of state, an approximate electron capture treatment during collapse, and a neutrino leakage scheme for the postbounce evolution.

Regarding BBH signals we consider the BBH waveform catalog of Mrou\'e et al~\cite{Mroue:2013} which includes the late inspiral, merger, and quasi-normal mode ringdown signals for 174 different models. Those waveforms have been computed using the Spectral Einstein Code ({\tt SpEC})~\cite{spec}. In addition we employ  the `R1'  BBH waveform computed by the GSFC group~\cite{baker:2007} and available at~\cite{astrogravs}. This waveform includes the late inspiral and merger phases of 
an irrotational BBH simulation performed with a grid-based numerical code, conformally flat initial data and the BSSN system of equations. The numerical code and techniques are hence different to those used by~\cite{Mroue:2013}.

\section{Dictionary generation and parameter evaluation}
\label{section:parameter_estimation}

\subsection{Dictionary generation}

We now turn to describe the process to generate a learned dictionary from the waveforms of both catalogs. The goal is 
to find the best set of dictionary parameters that produce the best denoising results. As  in~\cite{Torres:2014} we find that 
the results depend critically on the value of the regularization parameter $\lambda$ selected. The way we build the 
dictionaries for the burst and BBH catalogs is similar. We divide randomly in three groups both the 128 burst waveform 
signals of~\cite{Dimmelmeier:2008} and the first 100 BBH waveforms of~\cite{Mroue:2013}. Since the BBH signals 
of~\cite{Mroue:2013} are quite large, we do not use the entire BBH catalog in order to save computational 
resources. { Specifically, the BBH catalog covers binaries with total mass 20$M_{\odot}$ and mass ratios up to 1:8, 
and so does our dictionary.}  We then use in either case $80\%$ of the waveforms for training the dictionary, $15\%$ for 
validation of the method, i.e.~to search the best set of parameters, and the remaining $5\%$ to test the algorithm in 
different situations.

The numerically generated signals are embedded in non-white, Gaussian noise { corresponding to Advanced LIGO proposed broadband configuration,} provided by the LSC Algorithm Library Suite (LAL)~\cite{url:LAL}. The frequency ranges from 10 Hz to 8192 Hz (one-sided spectrum). First of all, we resample the waveforms of both catalogs to the Advanced LIGO/Virgo sampling rate of $16384$ Hz, zero padded to have the same length.  The corresponding signals are also shifted to be aligned with either the minimum peak in the case of bursts or with the maximum peak in the merger part for BBH signals. We select $2048$ samples around the corresponding alignment points to train the dictionary. With this length, the waveforms of the burst catalog fit completely in the window, while only the last cycles of the inspiral, merger and ringdown of the BBH waveforms are taken into account to perform the denoising. This late part of the BBH signal is arguably the most interesting part, hence deserving to be denoised best. Below we comment on the reason for this choice and on existing alternatives to also reconstruct the early inspiral part of the signal. 

To ensure the best conditions for the convergence of the algorithms and to avoid round-off errors, we also scale the amplitude of the validation signals of both catalogs so that their maximum value is set to unity. The values of the regularization parameter $\lambda$ we discuss in this section are hence determined by this normalization. Moreover, we scale each signal to achieve a specified value of the SNR, defined as
\begin{equation}
\text{SNR}=\sqrt{4\Delta t^2 \Delta f \sum_{k=1}^{N_f}\frac{|\tilde{h}(f_k)|^2}{S(f_k)}}~,
\label{SNR}
\end{equation}
where $\tilde{h}$ indicates the Fourier transform of signal strain $h$, $S$ is the power spectral density (PSD) of the noise, i.e.~the 
sensitivity curve of the detector, $f_k$ is each of the components of the frequency vector, $N_f$ is the number of positive frequencies, 
and $\Delta t$ and $\Delta f$ are the time step and frequency step, respectively. 

The optimal value of the regularization parameter, 
$\lambda_{\rm opt}$, is defined to be the one which 
gives the best results according to a suitable metric function
applied to the denoised signal and the original one, measuring the
quality of the recovered signal. 
In our case we choose two estimators, namely the Mean Squared Error, 
\begin{equation}
\label{eq:MSE}
\text{MSE} = \frac{1}{n}\sum_{i=1}^{n}(\hat{Y}_i-Y_i)^2~,
\end{equation}
where $\hat{Y}$ and  $Y$ are the reconstructed and original signals, respectively, and $n$ is the number of samples, and the structural similarity (SSIM) index~\cite{Wang2004}, which deviates from the traditional measures of error because it takes into account the structural information. The
SSIM index varies between 0 (minimum similarity) and 1 (maximum similarity) and is defined as 
\begin{equation}
\rm{SSIM}(x,y) = \frac{(2\mu_x\mu_y + c_1)(2\sigma_{xy} + c_2)}{(\mu_x^2 + \mu_y^2 + c_1)(\sigma_x^2 + \sigma_y^2 + c_2)},
\label{eq:ssim}
\end{equation}
where  $c_1$ and $c_2$ are constants, $\mu_x$ ($\mu_y$) is the average of $x$ ($y$), $\sigma_x^2$ ($\sigma_y^2$) the variance of $x$ ($y$) and $ \sigma_{xy}$ the covariance of $x$ and $y$.
 \begin{figure*}
  \centering
    \includegraphics[width=85mm]{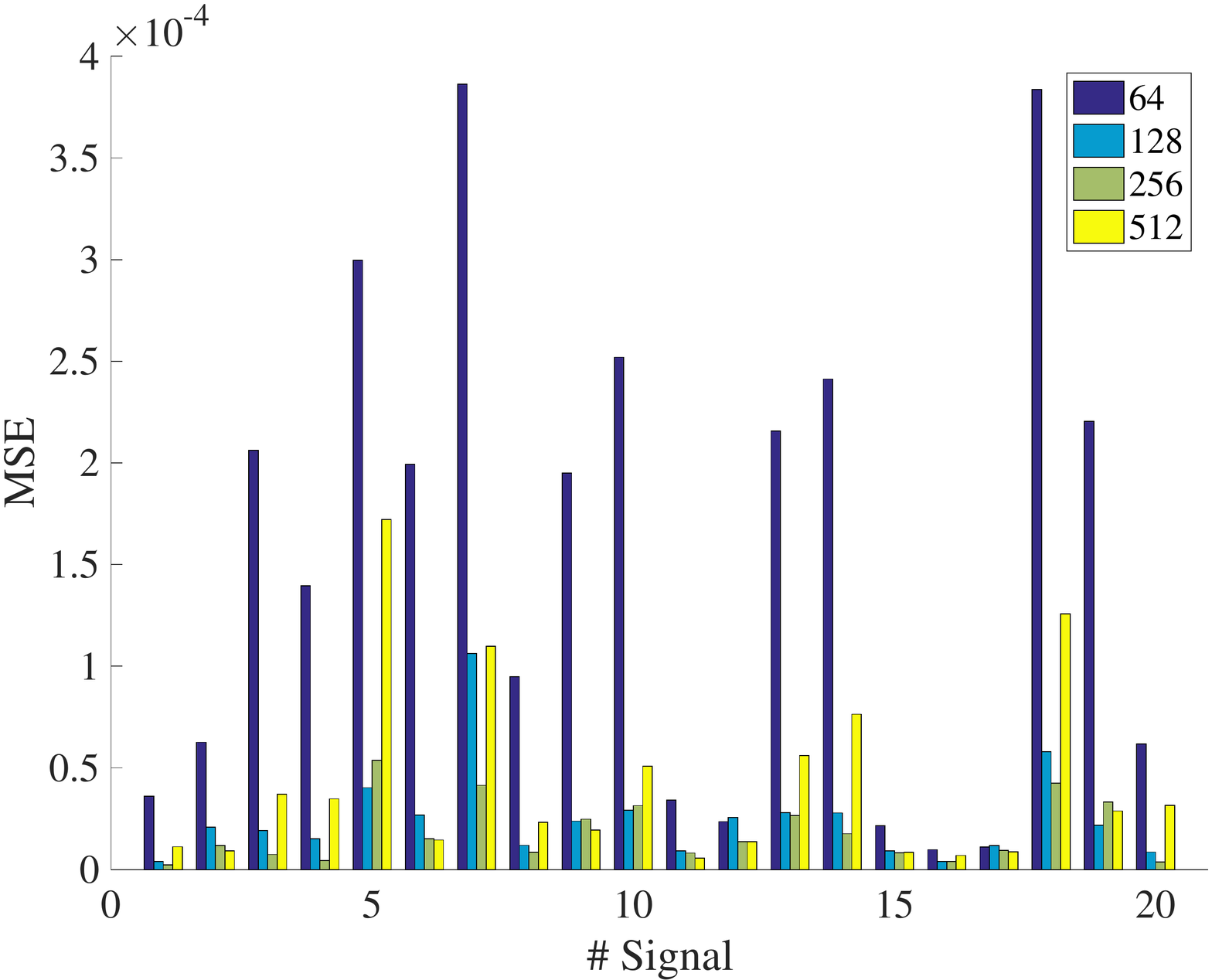}
    \hspace{0.2cm}
      \includegraphics[width=85mm]{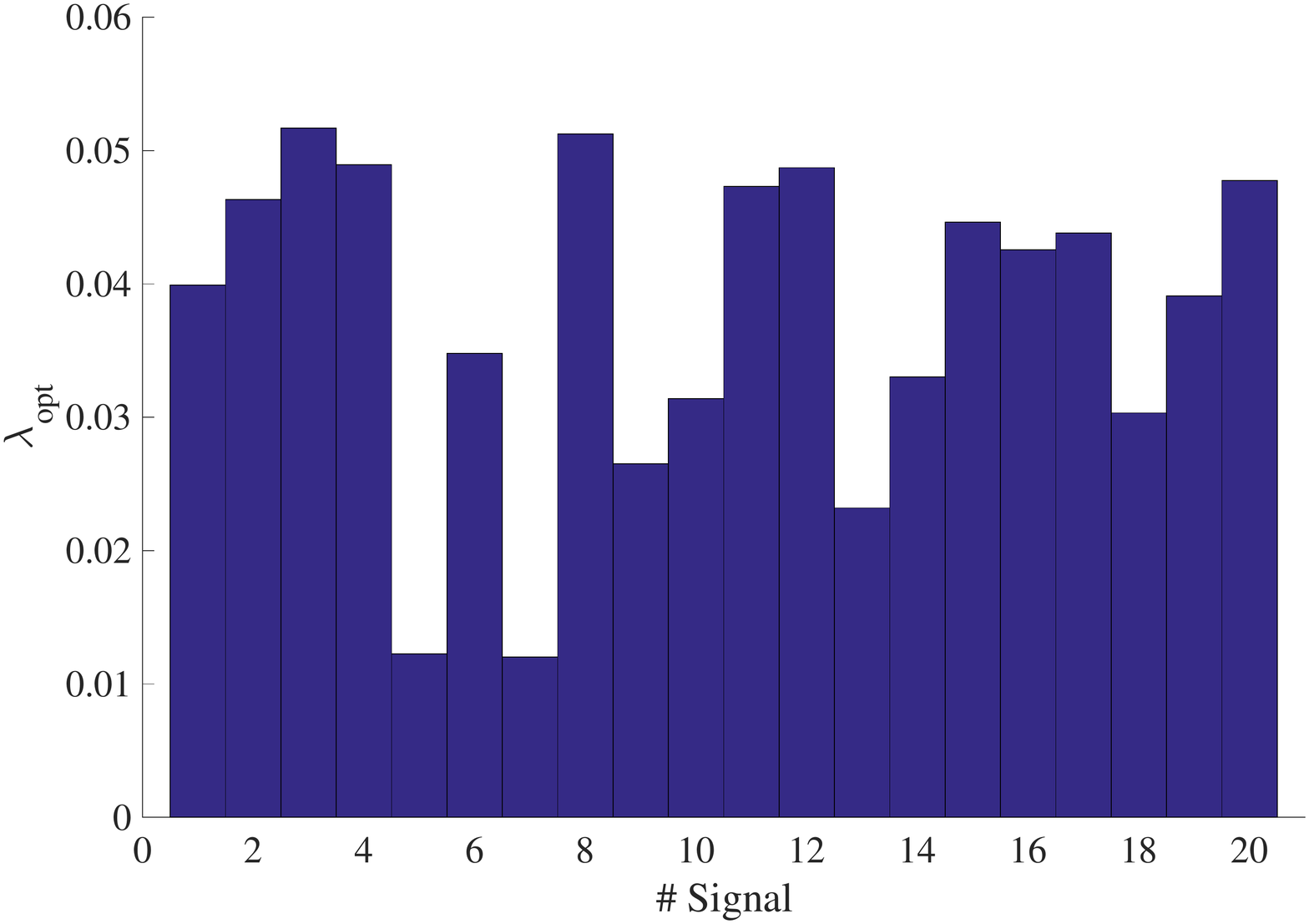}
  \caption{{\it Left}: Bar diagram of the MSE for all burst validation signals. Each color represents a different window length as indicated in the legend. {\it Right}: Bar diagram of the optimal value of $\lambda$ for all burst validation signals and a window length $n=256$. The mean value of $\lambda_{\rm opt}$ is $0.038\pm0.018$.}
   \label{fig:err_burst}
\end{figure*}

As mentioned before we use $80\%$ of the signals of each catalog 
to produce one dictionary per type of signal. To do the learning, we select $30000$ random patches (the starting sample is random) of a selected length, which is a parameter to be estimated. The patches are selected uniformly from all the learning waveforms of each catalog. Then, we select the $p$ patches with the highest energy, defined as the square of the $L^2$-norm of each patch. After that, we solve problem (\ref{eq:dict_learning}) using a block-coordinate descent method. This step is done modifying the code developed by \cite{Peyre:2011}. Fig.~\ref{fig:dictionary} shows a small representation of the atoms of both dictionaries. 

In addition to the search of $\lambda_{\rm opt}$ we must decide 
the best values for the size of the dictionary, i.e.~the number of atoms and their length. To this aim we calculate the MSE for the reconstructed signals obtained using dictionaries of different sizes. In each case, the value $\lambda_{\rm{opt}}$ will be the corresponding value that minimizes the MSE. For this task we use the validation set of signals of the dictionaries and set the SNR to $20$. As the length of the atoms is always shorter than the length of the validation signals, we do the denoising with a sliding window with an overlap of $n-2$ samples, where $n$ is the length of the window, which agrees with the length of the atoms. With this overlap, there are many samples that are repeated on different windows. These samples must be averaged to obtain the final reconstructed signals. Our initial tests show that the best reconstruction is achieved using TV-averaging (see~\cite{Marquina:2008}),

\begin{equation}
\label{eq:TV-averaging}
s= \frac{\sum_{j=1}^{p}(f_j\text{TV}_j)}{\sum_{j=1}^{p}\text{TV}_j}~,
\end{equation}
where $f_j$ corresponds to the current patch and 
$\text{TV}_j = \sum |\nabla f_j|$ is the Total-Variation norm of that patch. 

\subsection{Parameter evaluation}

We calculate the MSE for the validation set of burst signals with window lengths $l=64$, 128, 256, and 512. The results are shown 
in the left panel of Fig.~\ref{fig:err_burst}. Each vertical bar represents the value of
the MSE for each atom length. The figure shows that the largest value of the MSE is achieved for a length of $64$. This is due to the fact that if the atoms are too short the reconstructed signal is more oscillatory due to the noise. This effect
can be corrected using larger lengths. However, the larger the length of the atoms the more difficult to recover the smallest oscillations of the original signal. This is the reason why the MSE actually grows for the larger window length analyzed ($512$ samples). While this is a generic trend, it is nevertheless still possible that the longest window may work better for specific signals (e.g.~signals \#6 or \#11 in Fig.~\ref{fig:err_burst}). However, in general the best results correspond to a length of $256$ samples.

The right panel of Fig.~\ref{fig:err_burst} displays the values of $\lambda_{\rm{opt}}$, i.e.~the value of the regularization parameter that minimizes the MSE value. It has been obtained for a fixed window length $l=256$. This figure reveals that the values of $\lambda$ are bounded between $0.01$ and $0.06$. Therefore, not all values of $\lambda$ are possible and selecting the mean value $\lambda_{\rm{opt}}= 0.03$ will produce, on average, a good reconstruction for all burst signals. Nevertheless, fine-tuning this parameter can improve the results in specific cases. 

 \begin{figure*}
  \centering
    \includegraphics[width=85mm]{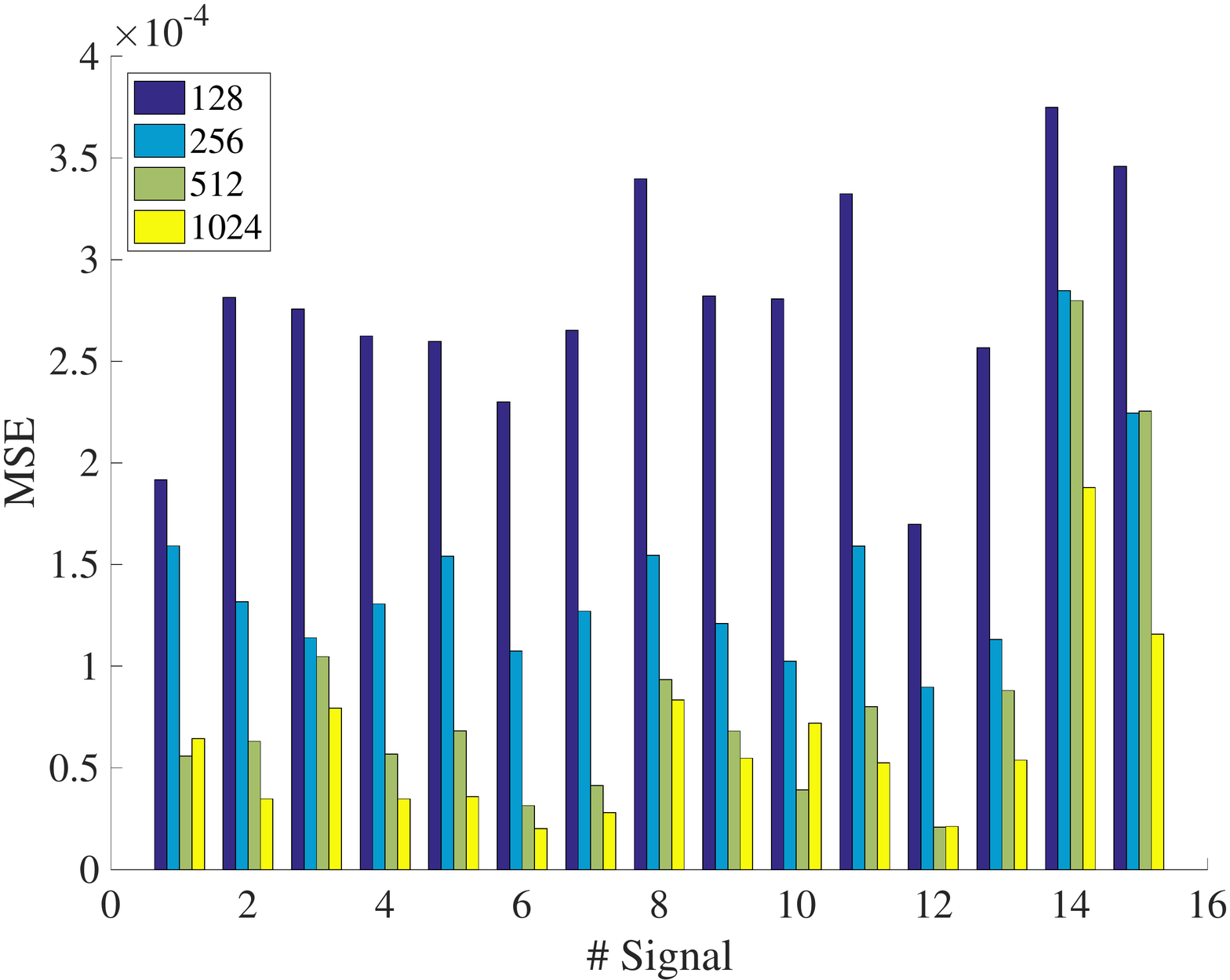}
    \hspace{0.2cm}
    \includegraphics[width=85mm]{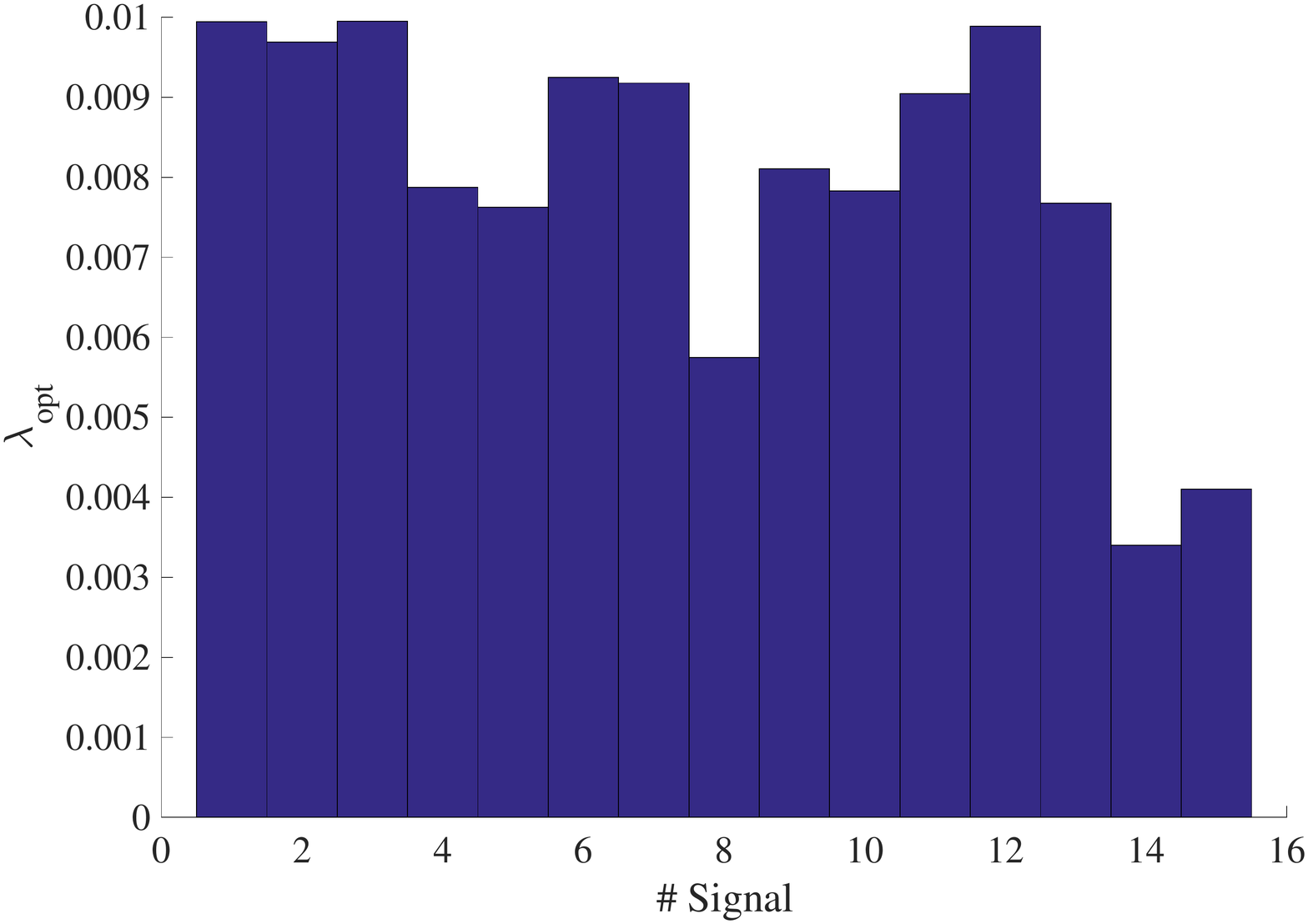}
  \caption{{\it Left}: Bar diagram of the MSE for all the BBH validation signals. Each color represents a different window length as shown in the legend.
  {\it Right}: Bar diagram of the optimal value of $\lambda$ for all the BBH validation signals and a window of length $1024$. The mean value of $\lambda_{\rm{opt}}$ is $0.008\pm0.002$.}
  \label{fig:bbh_error}
\end{figure*}

We next carry out the same analysis for the case of BBH signals. As BBH waveforms are totally different to burst signals, the choices just discussed for bursts would not lead to satisfactory results if applied blindly to the BBH catalog.
Contrary to burst signals, BBH waveforms are significantly longer, therefore we need to increase the length of the atoms. The values obtained for the MSE for the BBH catalog are shown in the left panel of Fig.~\ref{fig:bbh_error} and correspond to atom lengths that comprise from $128$ to $1024$ samples. The case of $64$ samples is not shown in the figure because the corresponding value of the MSE 
is much larger. As for the case of burst signals, Fig.~\ref{fig:bbh_error} shows that the MSE decreases with the window length in most cases. Therefore, to denoise BBH waveform signals we select the length of $1024$ samples as it produces the best results. 

The corresponding results for the value of $\lambda_{\rm{opt}}$ for BBH are shown in the right panel of Fig.~\ref{fig:bbh_error}. Again, the values are restricted to a small interval between $0.002$ and $0.012$.  As we show below, using the mean value, ${\lambda_{\rm{opt}}}= 0.008$, yields to satisfactory denoising results in most cases.

A similar study is required to determine how the results depend on the {\it number} of atoms of the dictionary $p$. In general,  the larger the dictionary the better the results, but at a higher computational cost. Therefore, setting the size of the dictionary is often a trade-off between results quality and efficiency. To evaluate an optimal value for the number of atoms we carry out tests with the two catalogs using values from $p\in\{300, 500, 1000\}$ in the case of bursts with $n=256$ atom length and from $p\in\{1100,2000,2500\}$ in the case of BBH with $n=1024$ atom length. We find that using $500$ and $1100$ atoms for bursts and BBH, respectively, is a valid compromise as it produces good results at a reasonable computational cost. However, if computational resources are not an issue, there is no reason not to use larger dictionaries. For the two catalogs, the value of $\lambda_{\rm{opt}}$ for $p=500$ and $p=1100$ atoms are bounded in a similar interval than shown before. 

\section{Tests and results}
\label{section:application}
 
\subsection{No signal}

\begin{figure}
  \centering
    \includegraphics[width=80mm]{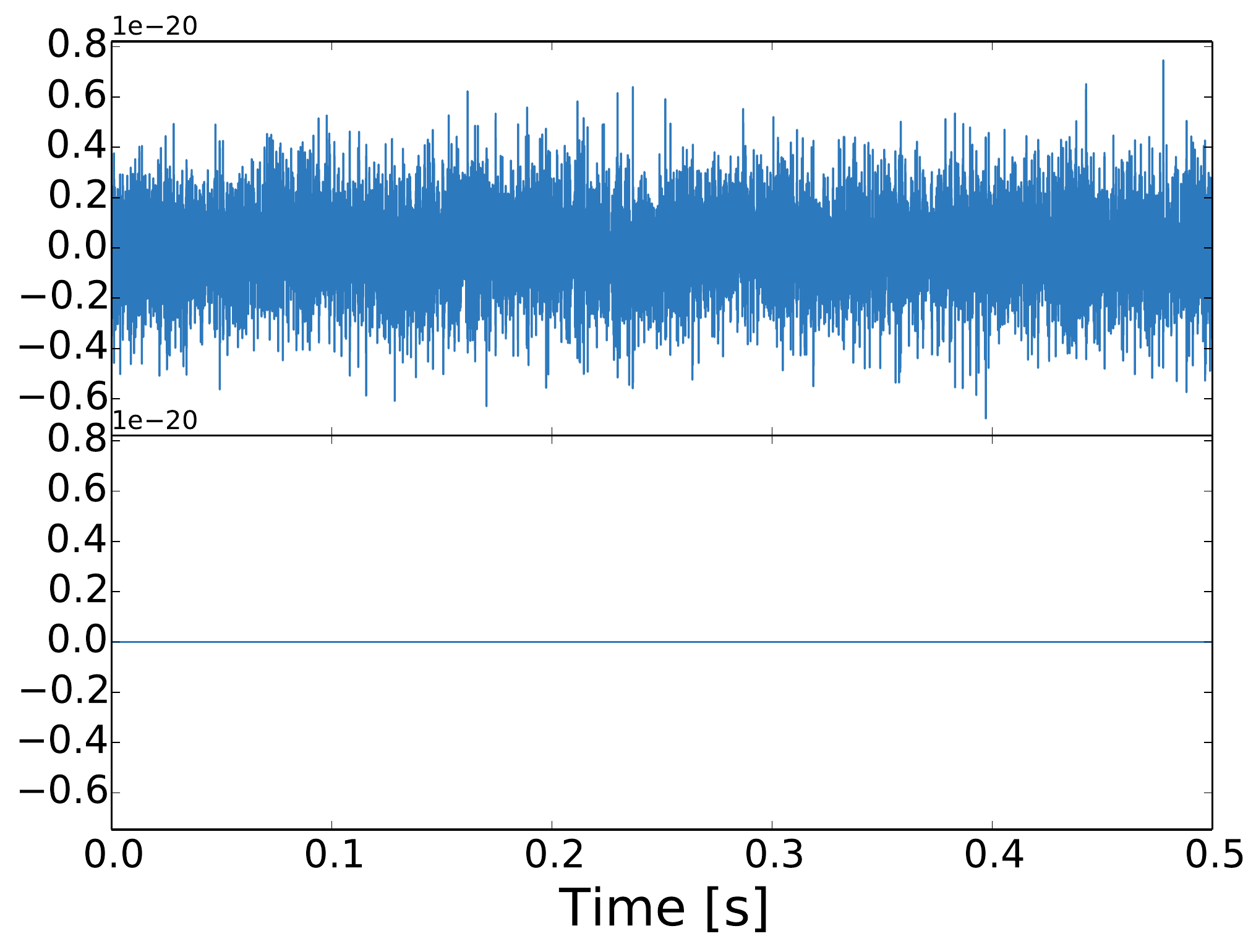}
  \caption{Denoising with no signal embedded into Gaussian noise. The upper panel shows the original noisy signal while the lower panel shows the { ideal} result of the denosing, i.e.~a zero amplitude signal.}
  \label{fig:no_signal}
\end{figure}

The first test consists in studying the performance of the method when there is no signal inside the data set. The goal of this test is to 
check if in the absence of signal the dictionary produces spurious signals due to noise. The result of this test is shown in Fig.~\ref{fig:no_signal}. A stream of $0.5$ s of pure non-white Gaussian noise (upper panel) is denoised using the generic value of $\lambda_{\rm opt}$ corresponding to burst signals, i.e.~$\lambda_{\rm opt}=0.03$.  One can see that the resulting signal has zero amplitude
throughout the frame (lower panel) for this specific value of $\lambda$. This is the { ideal}  behavior of the algorithm in order to avoid false detections due to noise.


{ We next repeat this test for 200 independent realizations of noise (following the procedure outlined in Appendix A 
of~\cite{Torres:2014}) 
to check if this behavior remains the same irrespective of the
noise realization. For our specific value of $\lambda$ we find 26 false reconstructions due to noise fluctuations.}  We note however 
that the smaller the $\lambda$ the more coefficients of the representation become nonzero and { more} structures due to noise may appear. { In contrast, a large value of $\lambda$ will reduce the ratio of false reconstructions, even though a true GW signal with low SNR could be missed. For instance, for $\lambda=0.045$ we only obtain one false reconstruction.} 

{ The results reported in this section are illustrative of the typical response of the LASSO algorithm on $\lambda$.
A comprehensive statistical study of the dependence of the number of false reconstructions and signal misses on the 
parameters of the method, i.e.~value of $\lambda$, type of signal injection, SNR, and noise realization, deserves further 
analysis. We also note that this is a fairly simple test because the noise is purely Gaussian. In a more realistic scenario, 
the presence of instrumental glitches in the detector data~\cite{Powell:2015,Powell:2016} could produce false reconstructions.}

\subsection{Signals from the catalogs}
%
\begin{figure*}
  \centering
    \includegraphics[width=85mm]{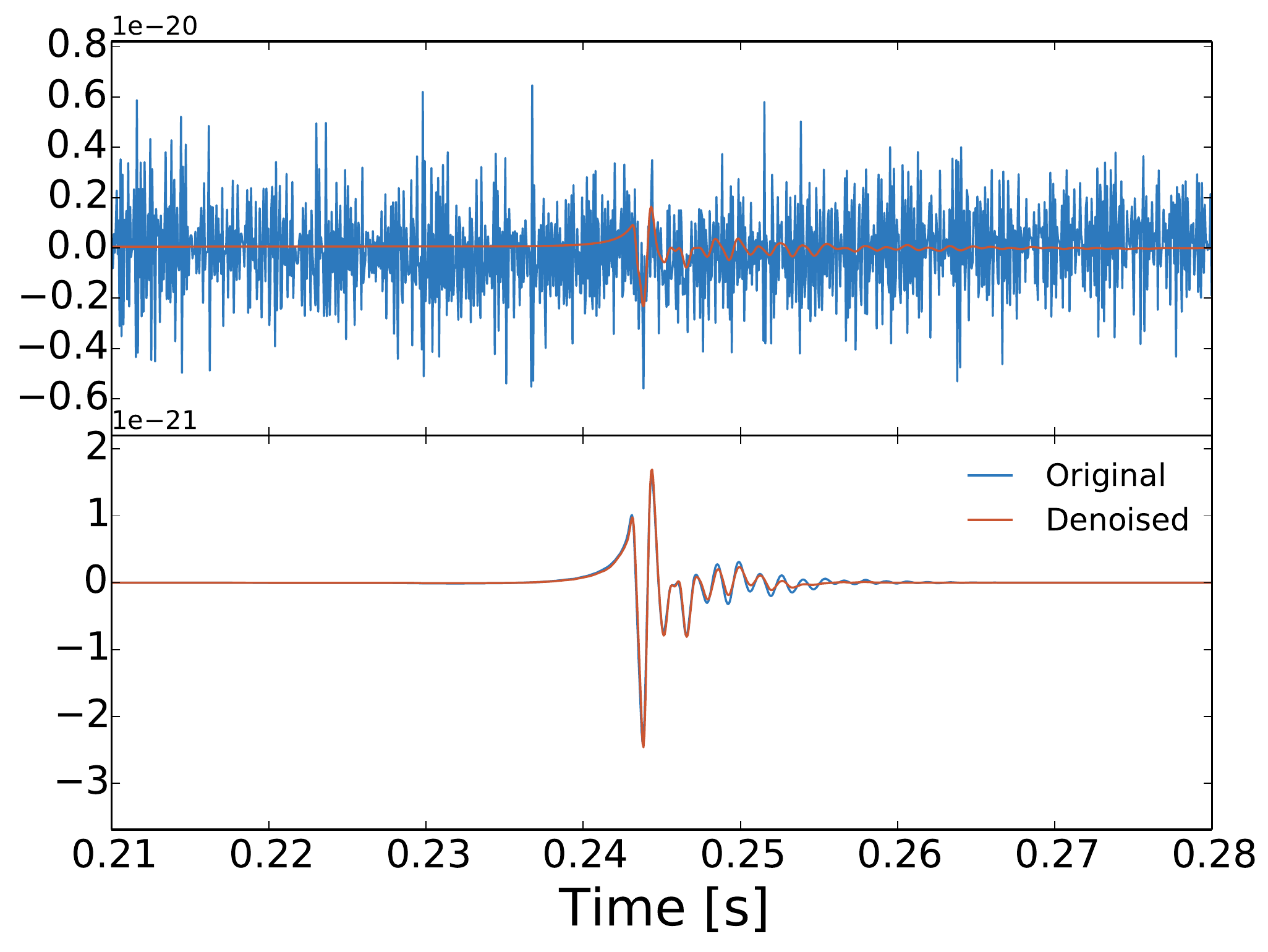}
        \includegraphics[width=85mm]{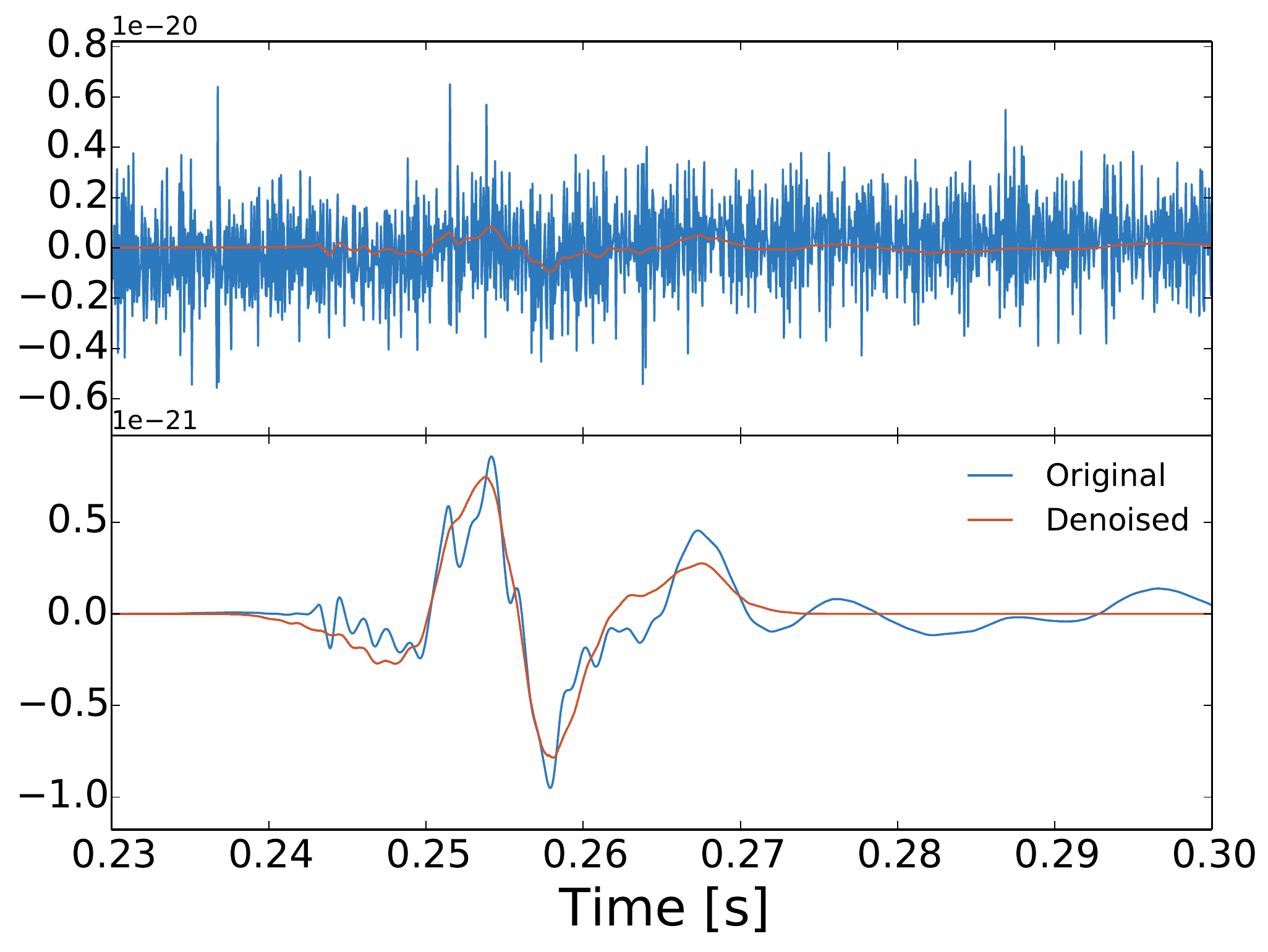}
  \caption{Denoising of signal \#1 (left) and \#6 (right) from the group of test signals of the burst catalog. The time of arrival is random and the SNR is 20. Upper panels: noisy signals (blue) superimposed with the original numerical relativity waveforms (red). Lower panels: comparison between the denoised signals (red) with the original ones (blue). { The MSE and SSIM values are  $0.018\times 10^{-3} $ and $0.98$ for the signal on the right panel and $0.271  \times 10^{-3} $  and $0.67$ for the signal on the right panel. }}
  \label{fig:burst_denoising}
\end{figure*}
Next we study how the method works when applied to the eight test waveform signals of the burst catalog in a long data frame. 
{ In the figures for this and the following tests, we use the same noise realization to compare the results on an equal footing. Correspondingly, in the tables reported in this section, we present results obtained with 20 different noise realizations 
to find out how the reconstruction is affected by noise fluctuations.}
The signals are embedded in Gaussian noise with a SNR of $20$. The time of arrival is fixed and it is the same for all the signals. The value of the regularization parameter is set to $\lambda=0.03$ and remains the same value for all the tests of this section. Although this value is not the optimal one, i.e.~the one which produces the best results for a given signal, our goal is to determine if it is possible to recover the signal with a generic value of $\lambda$. This approach may be closer to what occurs in a realistic situation, where no information on the signal is available a priori. 

The quality of the denoising is measured using the MSE and the SSIM metric functions, and is reported in Table \ref{Table:Burst} for all test signals. { This table shows the maximum and the minimum values for both MSE and SSIM for 20 independent noise realizations. 
We recall that the results depend on the value of $\lambda$. Each signal embedded in different noise realizations is a new scenario, and the best results 
will be obtained with the optimal value of $\lambda$ for each case. With SNR 20 and $\lambda=0.03$ the relative variations are not too large (the highest variation in SSIM is 14\% for signal \#6).
Therefore, at this SNR, the 
reconstruction is not very affected by noise fluctuations. }
Fig.~\ref{fig:burst_denoising} shows the results for only two signals of the catalog, namely those which yield the best (signal \#1; left panel) and the worst (signal \#6; right panel) denoising results, respectively { (for the chosen value of $\lambda$ and noise realization)}. The figure displays 
the comparison of the two original noisy signals (upper panels) with the recovered ones (lower panels). 
Concerning the signal on the left panel our method can accurately recover the distinctive positive and negative peaks associated with the hydrodynamical bounce that follows the collapse of the inner iron core of the star once the equation of state stiffens and the central density exceeds nuclear matter density. This is particularly clear for the peaks with the larger amplitudes, which are recovered properly. However, when the amplitude decreases (i.e.~in the part of the temporal evolution associated with the quasi-radial oscillations of the newly formed neutron star) the signal becomes weaker than the noise and, as a result, the method returns a zero amplitude signal.
It is also worth mentioning that in the part of the time series where the data are purely noise (no numerical relativity signal injected) the method returns a zero signal, as it should. 
The same behavior is seen for the signal displayed on the right panel of Fig.~\ref{fig:burst_denoising}, the dampened oscillations are weaker than the noise and the method sets their amplitude to zero. We note that signal \#6 is somewhat different from the common features of the dictionary. As a result, while the broad morphology is still captured to some extent, the overall result is poorer than for the signal on the left panel. Even so, we note that the results can be improved by changing slightly the value of $\lambda$ by adding more atoms to the dictionary. (We have checked that for $\lambda=0.026$ the MSE is  $0.1 \times 10 ^{-3}$ and the SSIM is $0.77$.) 

\begin{table}[t]
 \centering
  \caption{Values {(maximum - minimum)} of the MSE and SSIM error estimators for the eight burst signals we use as test signals { and 20 noise realizations}. Values are reported for
  both SNR 20 and 10. }
  \begin{tabular}{@{}cccccc@{}}
  \hline
   {Signal }    & \multicolumn{2}{c}{SNR 20}& &\multicolumn{2}{c}{SNR10} \\
                    & MSE ($\times10^{-3}$) & SSIM & & MSE ($\times10^{-3}$) & SSIM \\
 \hline	

\#1 & [0.033 - 0.015] & [0.97 - 0.93]&& [1.389 - 0.021]& [0.96 - 0.74] \\ 
\#2 & [0.124 - 0.030] & [0.95 - 0.85]&& [1.264 - 0.125]& [0.89 - 0.60] \\ 
\#3 & [0.066 - 0.040] & [0.93 - 0.88]&& [0.691- 0.073]& [0.89 - 0.74] \\ 
\#4 & [0.068 - 0.007] & [0.97 - 0.88]&& [0.684 - 0.014]& [0.95 - 0.74] \\ 
\#5 & [0.052 - 0.022] & [0.94 - 0.89]&& [1.335 - 0.041]& [0.90 - 0.53] \\ 
\#6 & [0.210 - 0.084] & [0.83 - 0.72]&& [0.861 - 0.205]& [0.72 - 0.51] \\ 
\#7 & [0.130 - 0.083] & [0.90 - 0.84]&& [2.350 - 0.103]& [0.88 - 0.43] \\ 
\#8 & [0.042 - 0.016] & [0.93 - 0.85]&& [0.594 - 0.026]& [0.91 - 0.74] \\ 
       		
\hline
\label{Table:Burst}
\end{tabular}

\end{table}

\begin{figure*}
  \centering
    \includegraphics[width=85mm]{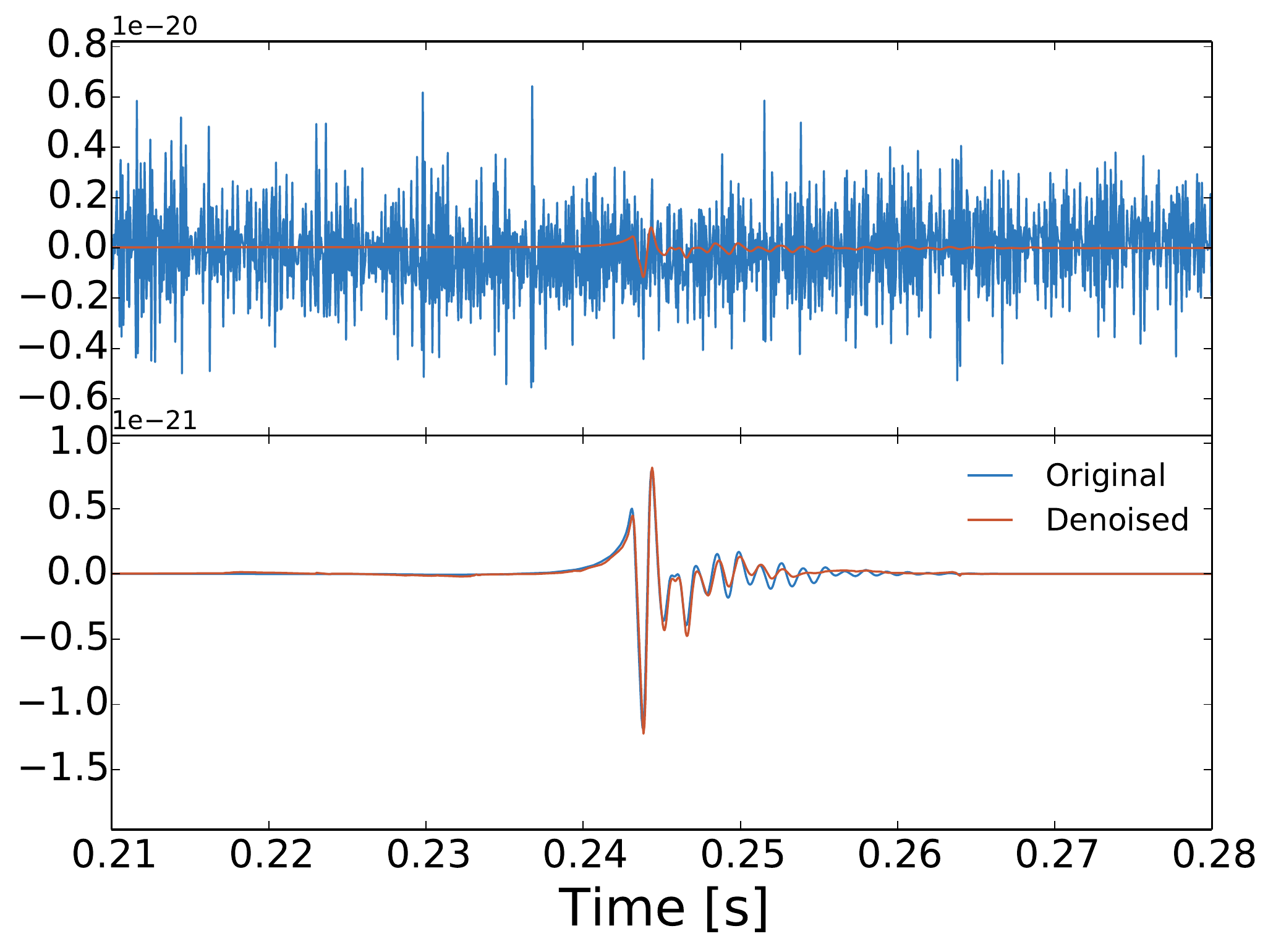}
    \includegraphics[width=85mm]{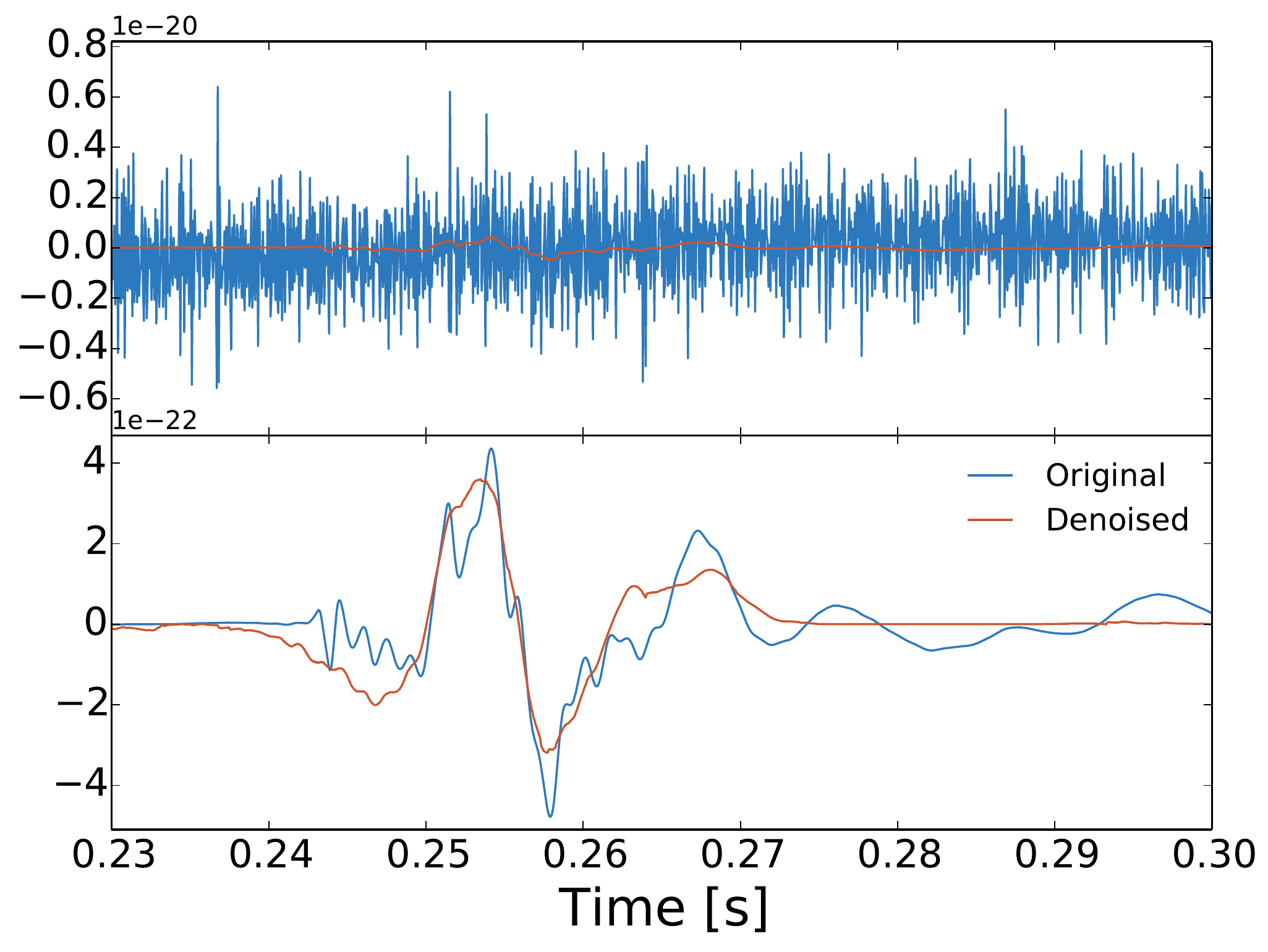}
  \caption{Same as Fig.~\ref{fig:burst_denoising} but with SNR $10$. { The MSE and SSIM values are  $0.081 \times 10^{-3} $ and $0.91$ for the signal on the right panel and $0.20  \times 10^{-3} $  and $0.71$ for the signal on the right panel. }}
  \label{fig:burst_signal10}
\end{figure*}

To find out the dependence of the procedure on the SNR we reduce its value from 20 to 10, keeping the same value of $\lambda$.
The results are displayed in Fig.~\ref{fig:burst_signal10} for the same signals \#1 and \#6 of the burst catalog. The results for all test signals and the corresponding { maximum and minimum} measures of the MSE and SSIM are also reported in Table \ref{Table:Burst}. Figure~\ref{fig:burst_signal10} shows that  for SNR 10 signal \#1 is still very well recovered and its most significant features can be reconstructed with relatively high accuracy. The MSE for this signal increases an order of magnitude and the SSIM decreases from $0.98$ to $0.91$, still reasonably high. For the worst possible case of the test waveforms, signal \#6,  Fig.~\ref{fig:burst_signal10} shows that it can still be distinguished from the noise. 

Comparing the results for all burst signals reported in Table~\ref{Table:Burst} for SNR 20 and SNR 10, we see that, in general, the values 
of the MSE (SSIM) increase (decrease) if the SNR decreases, except for the case of signal \#6. We recall that we are using the same value of $\lambda$ for all signals, and if it is not near the optimum value for any given signal, the results will not be good. In the case of signal \#6 the errors for SNR 20 are actually slightly worse than for SNR 10. { As the SNR decreases, the reconstruction is more affected by noise fluctuations and the difference between the maximum and minimum values of the quality indicators increases. As 
we have mentioned before, other values of $\lambda$ could improve the results in each case.}

\begin{figure}
  \centering
    \includegraphics[width=85mm]{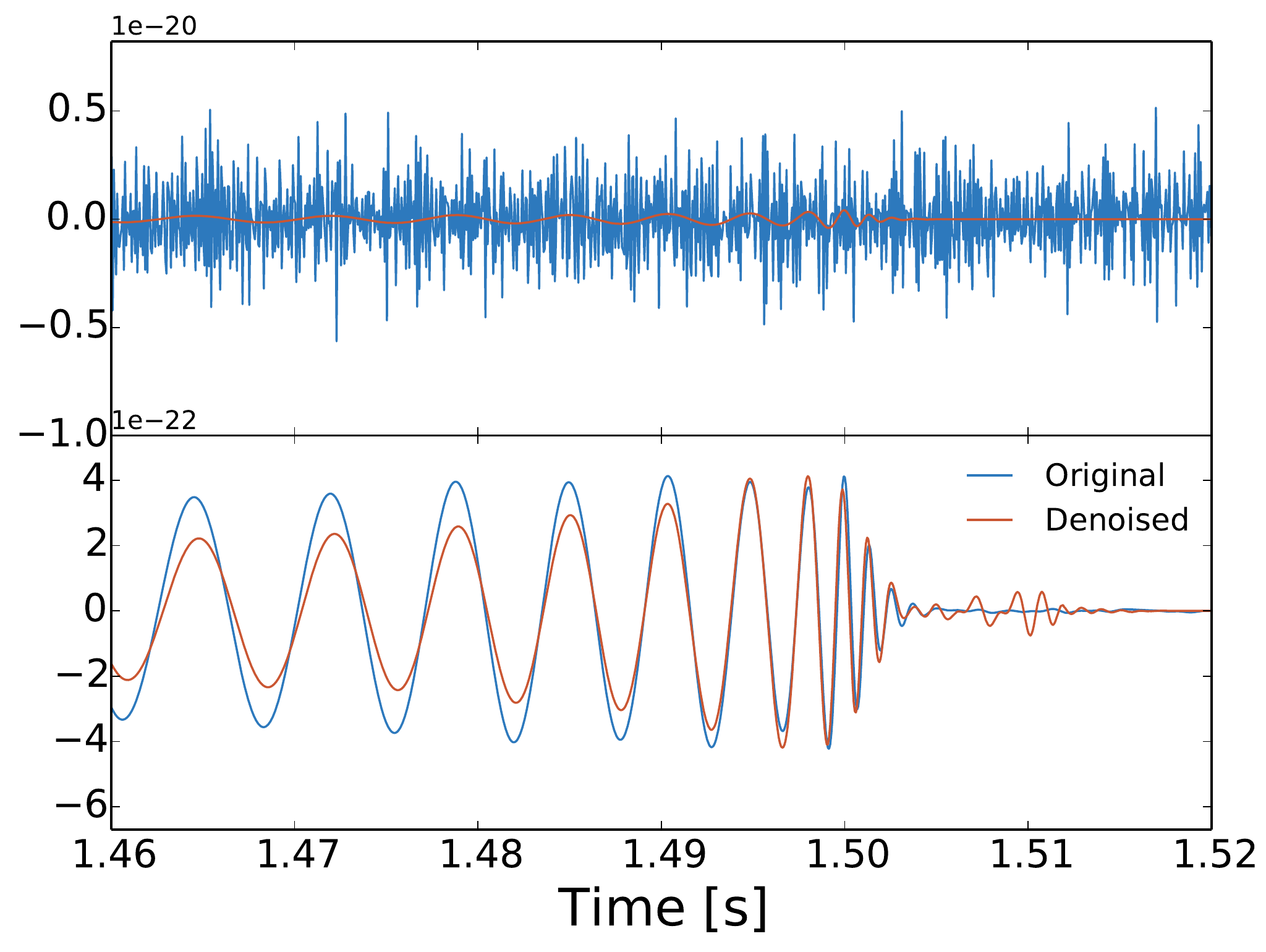}
  \caption{Denoising of the test signal \#2 taken from the BBH catalog. The SNR is set to $20$ in a $2$ s frame. The value of $\lambda$ used is $0.01$ with TV averaging. { The values of MSE and SSIM are 
  $0.031\times 10^{-3}$ and $0.86$ respecrespectively.}}
  \label{fig:bbh_signal}
\end{figure}

We turn next to test the results from the BBH catalog. As these signals are much longer than burst signals, the total segment of data has a length of $2$ s, in order to allow to change the time of arrival. In this case, it is set as the time where the merger is produced, randomly.  The denoising results are reported in Table \ref{Table:BBH} for all BBH signals and displayed in Fig.~\ref{fig:bbh_signal} for a representative signal (\#2). As the figure shows, the signal is properly denoised during its three distinctive parts, the inspiral, the merger, and the ringdown. 
In particular, the phase of the signal is well captured and the main, yet small, differences between the original and the denoised signal appear in the amplitude. We note that the actual signal is significantly longer than the zoom shown in this figure. The initial part of the signal, the inspiral phase with low frequencies, is not recovered because, as mentioned before, the dictionary is specifically designed to recover the merger part. The most striking incorrect feature of Fig.~\ref{fig:bbh_signal} is the presence of spurious oscillations visible after the ringdown. This is due again to the selection of $\lambda$. While using a larger value would remove these oscillations it is also possible that the amplitudes of the merger and ringdown parts of the signal could be cut down. The corresponding MSE and SSIM measures are reported in Table \ref{Table:BBH} for both SNR 20 and 10. As for the case of burst waveforms, for BBH signals the values of the MSE (SSIM) also increase (decrease) as the SNR decreases, as expected.

\begin{table}
 \centering
  \caption{Values {(maximum - minimum)}  of the MSE and SSIM error estimators for the four BBH signals we use as test signals { and 20 noise realizations}. Values are reported for both SNR 20 and 10.}
  \begin{tabular}{@{}cccccc@{}}
  \hline
   {Signal}    & \multicolumn{2}{c}{SNR 20}& &\multicolumn{2}{c}{SNR 10} \\
         &  MSE ($\times10^{-3}$)  & SSIM  & &MSE ($\times10^{-3}$) & SSIM \\
        \hline
\#1 & [0.025 - 0.019] & [0.89 - 0.86]&& [0.084 - 0.027]& [0.87 - 0.76] \\ 
\#2 & [0.060 - 0.027] & [0.86 - 0.79]&& [0.104 - 0.039]& [0.83 - 0.66] \\ 
\#3 & [0.029 - 0.020] & [0.88 - 0.87]&& [0.101 - 0.032]& [0.86 - 0.74] \\ 
\#4 & [0.034 - 0.019] & [0.89 - 0.86]&& [0.101 - 0.025]& [0.87 - 0.74] \\ 
       		
\hline
\label{Table:BBH}
\end{tabular}

\end{table}
%
\subsection{Signals not included in the catalogs}

In a realistic scenario, the gravitational wave signal will be unknown and it will be contaminated by several sources of noise. To test the code in a more unidealized setting we select in this section signals with similar broad morphology to those of the dictionaries but
generated in a different way (e.g.~employing different numerical codes or input physics). While such a situation is still simple, since it involves simulated Gaussian noise without glitches, it is nonetheless more realistic because, contrary to the cases analyzed before, the signals are now different from those of the catalogs from which the dictionaries are generated. 

\begin{figure}
  \centering
    \includegraphics[width=85mm]{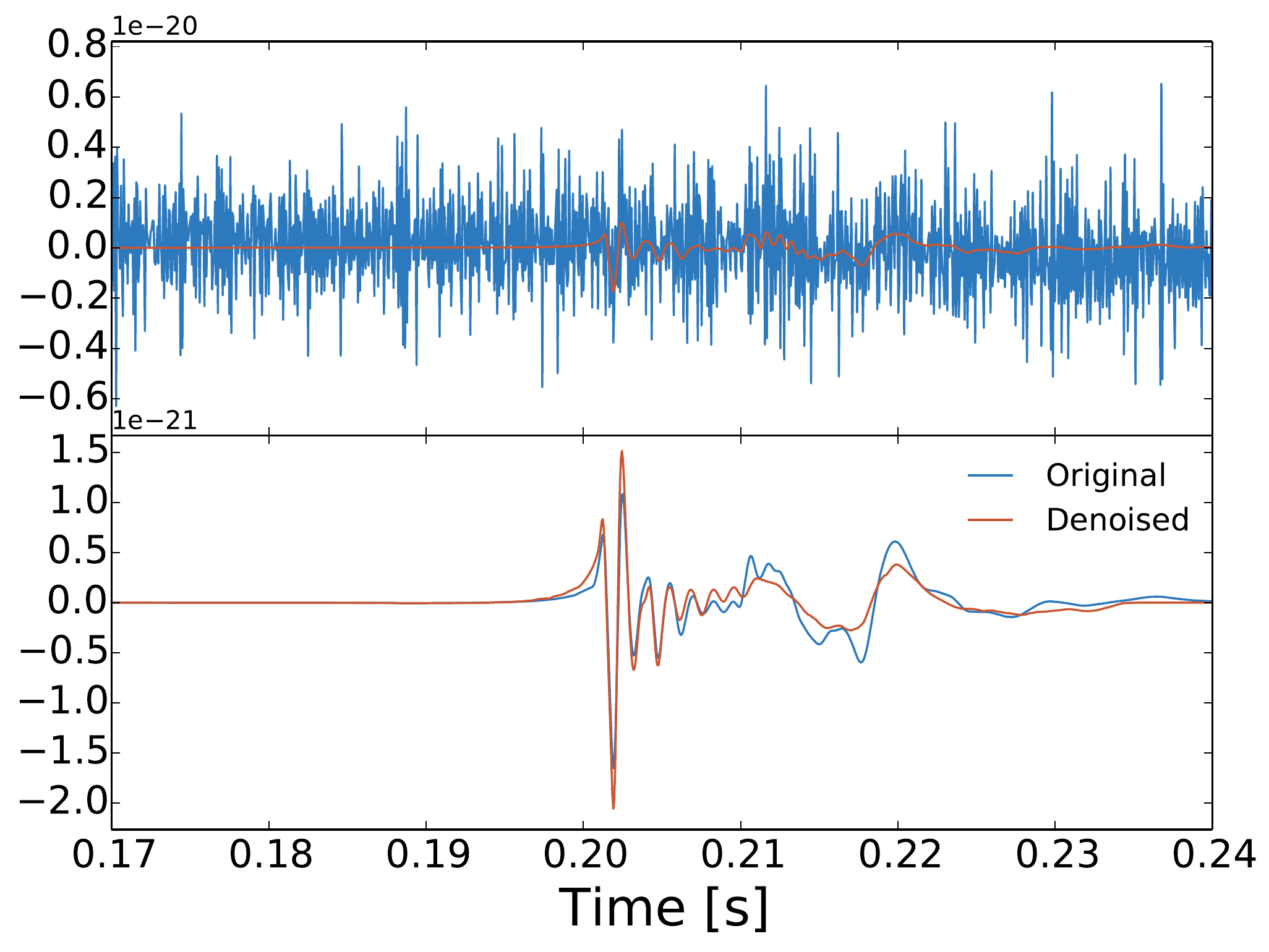}
  \caption{Denoising of a burst signal from the core collapse catalog of~\cite{abdikamalov14} using a dictionary generated from a different catalog~\cite{Dimmelmeier:2008}. The arrival time is random, the SNR is $20$ and $\lambda$ = 0.03.}
  \label{fig:abd_signal} 
\end{figure}

We first consider a burst signal from a core collapse catalog generated by~\cite{abdikamalov14}. We select signal \#1 from this catalog, embed it into Gaussian noise with a SNR 20, and proceed to denoise it employing our burst dictionary. The results of the denoising are displayed in Fig.~\ref{fig:abd_signal}. This figure shows that the positive and negative peaks associated with core bounce are well recovered. The values of the MSE and SSIM error estimates are respectively $3.78\times10^{-5}$ and $0.96$. Notwithstanding some characteristics of the signal are lost, the signal can nevertheless be clearly distinguished from the noise and the main features are well recovered. 

\begin{figure}
  \centering
    \includegraphics[width=85mm]{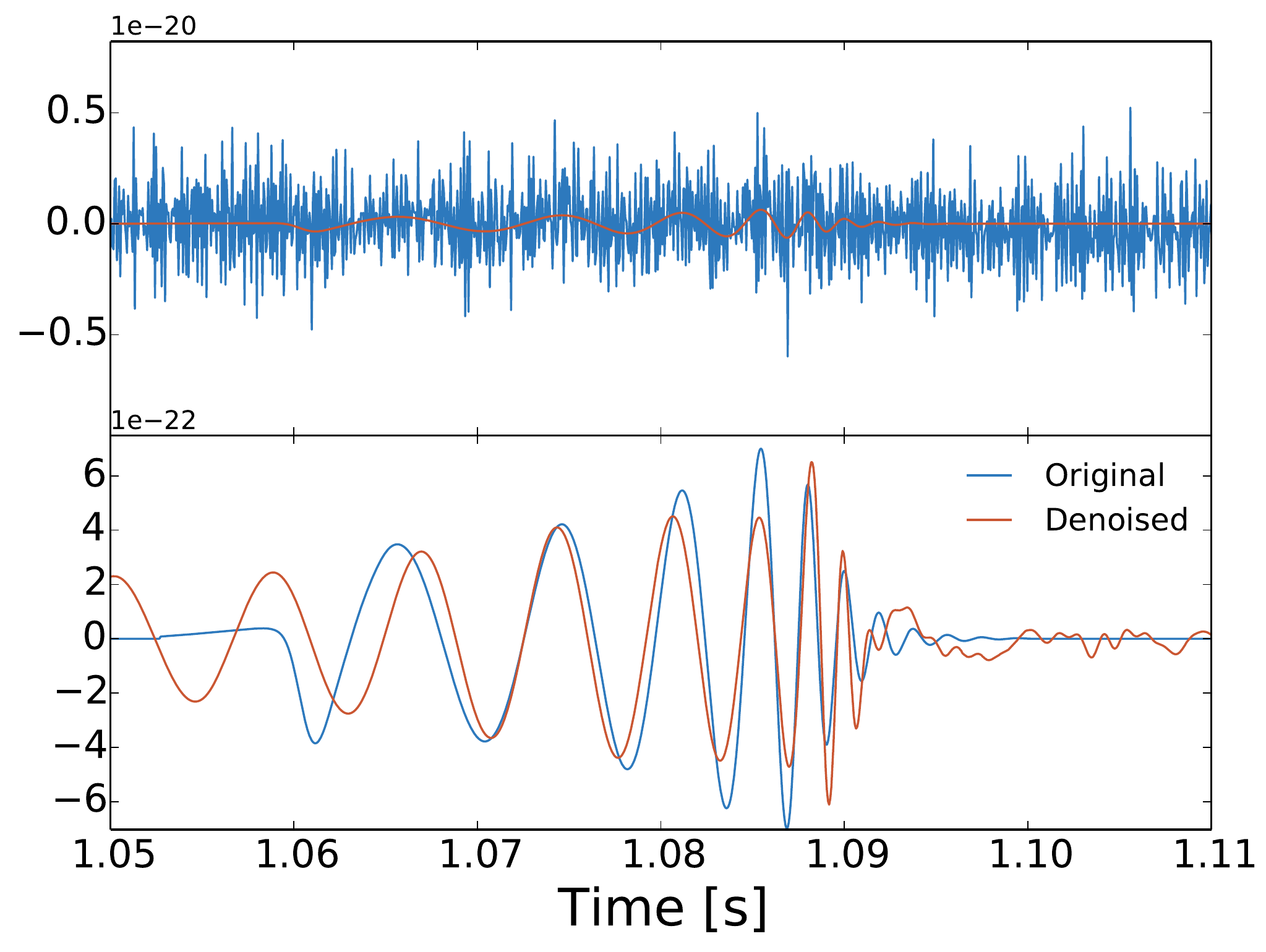}
  \caption{Denoising of the `R1' BBH signal computed by the GSFC group~\cite{baker:2007} using a dictionary trained with signals from a different catalog~\cite{Mroue:2013}. The SNR is set to $20$ and the time of arrival is random in a $2$ s frame. The value of $\lambda$ used is $0.09$ with TV averaging.}
  \label{fig:baker_signal}
\end{figure}

We can carry out a  similar study for our BBH dictionary. As mentioned before, we select the BBH signal `R1'  from~\cite{baker:2007}. The result of the denoising is displayed in Fig.~\ref{fig:baker_signal} which shows that the reconstruction is much less accurate than in the case of the BBH test signals discussed before. We obtain  $\rm{MSE}=5.79\times10^{-4}$ and $\rm{SSIM}=0.55$, values which indicate a poor reconstruction. The merger is not correctly recovered and the recontruction introduces a phase shift. We must recall once again that we are using a generic value of $\lambda$ and therefore the result could be significantly improved by choosing a more suitable value. However, the goal of this test is not to obtain the best result possible but to assess our procedure in a scenario where the incoming signal is unknown and differs from those used to train the dictionary. 

%
\section{Complementary Tests}
\label{section:other}

\subsection{Iterative denoising}

\begin{figure}
  \centering
    \includegraphics[width=85mm]{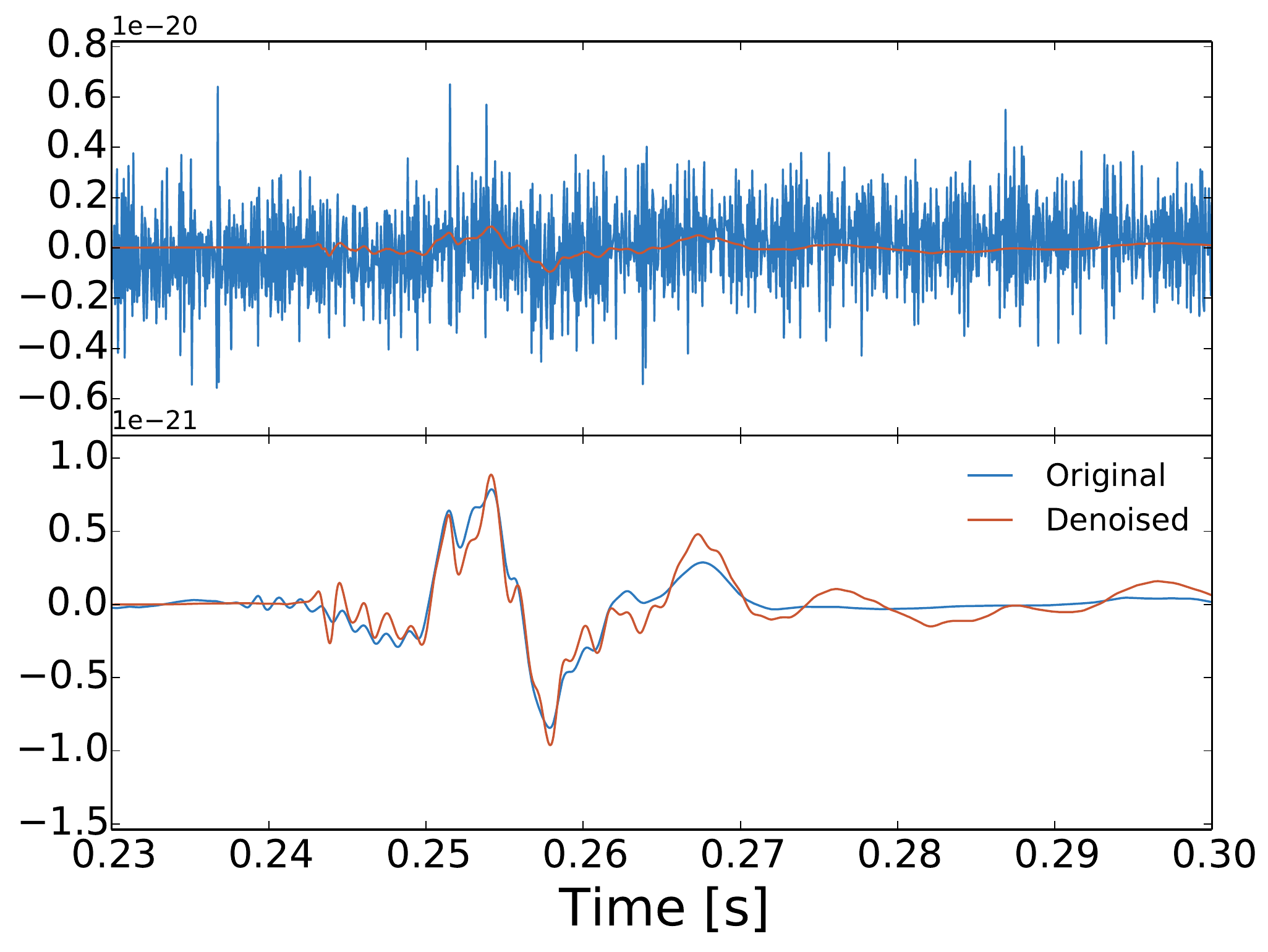}
  \caption{Denoising of signal \#6 of the burst catalog employing an iterative procedure with only 2 iterations. We choose $\lambda=0.01$ and SNR 20.}
  \label{fig:iterative}
\end{figure}

The next situation we consider involves a simple direct extension of the method, namely using the denoising procedure in an iterative way. 
By removing noise iteratively we find that the small amplitude oscillations of the signals are recovered better than using one single iteration. 
In this approach we use a generic, low value of $\lambda$, which only removes a small amount of noise in every iteration.
We apply this iterative approach to burst signal \#6 with SNR 20. The results are shown in Fig.~\ref{fig:iterative} (to be compared with
the right panel of Fig.~\ref{fig:burst_denoising}). For this SNR we find that typically only 2 or 3 iterations 
suffice to recover the small amplitude oscillations of the signal and improve the results. For this test we obtain 
MSE $=0.079\times10^{-3}$ and SSIM $=0.81$. These values are considerable better than those reported in Table~\ref{Table:Burst}
for the case of a single iteration. 

\subsection{Combination of signals}

\begin{figure}[t]
  \centering
    \includegraphics[width=85mm]{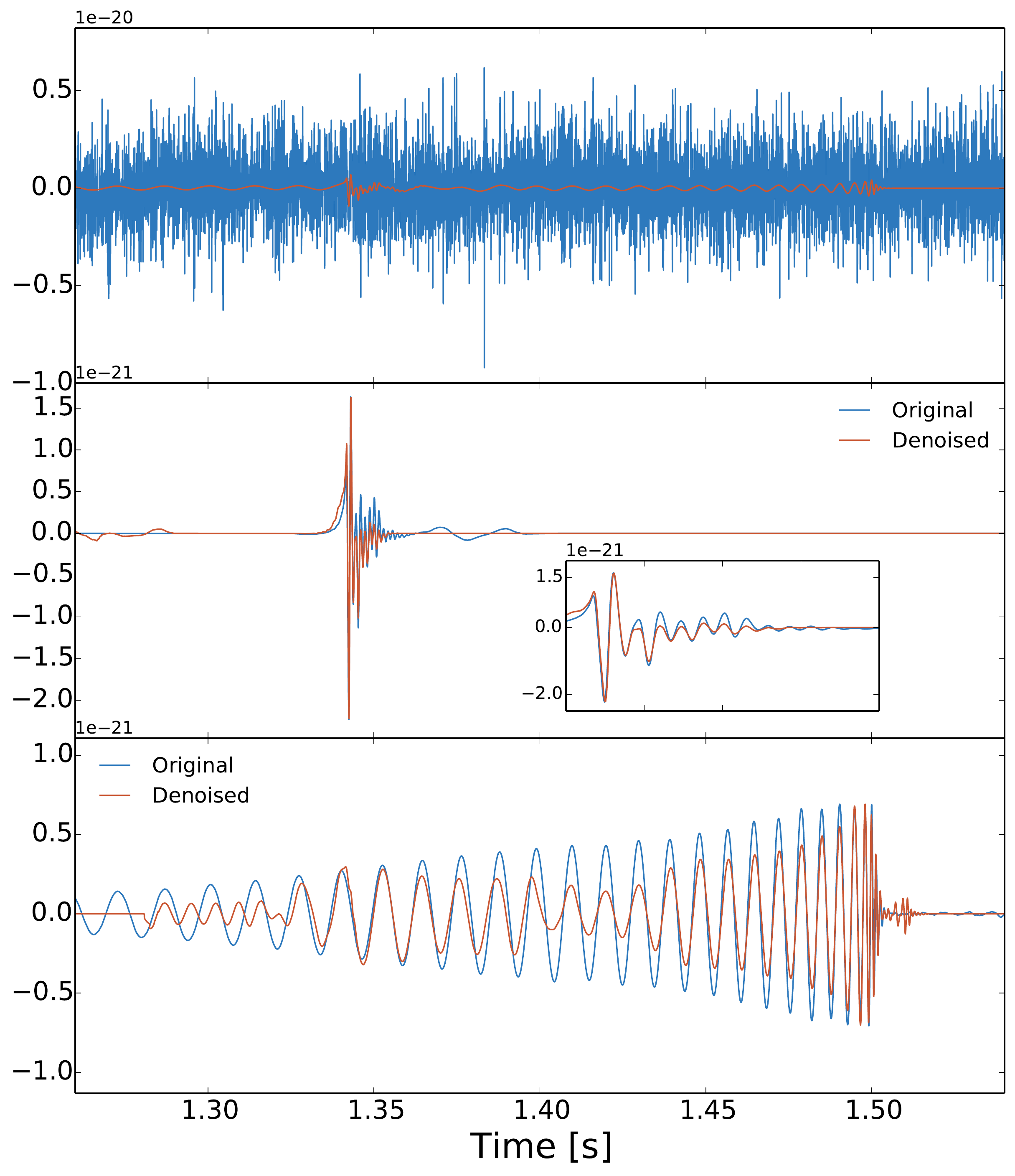}
  \caption{Denoising of a test signal composed by a combination of burst signal \#5 and BBH signal \#2. The individual signals are
  independently recovered when using the appropriate dictionary in a standalone way, as shown in the middle and bottom panels.}
  \label{fig:join_signals}
\end{figure}

For our next test we assess our procedure when the signal to denoise is a combination of two different signals. The goal 
of this test if to check the performance of our dictionaries when dealing with signals different from the type they are designed 
for. To do that, we build a test signal which is a combination of a burst and a BBH, both with SNR 20. We apply the algorithm 
using both dictionaries independently. The results of this test are shown in Fig.~\ref{fig:join_signals}. The upper panel shows 
the original test signal (red line) embedded in Gaussian noise. The burst is located around $t\sim 1.34$ s while the merger 
of the BBH signal is visible at $t\sim 1.50$ s. The middle panel shows the reconstruction (red line) using only the burst 
dictionary with a value of $\lambda=0.03$. The inset of this panel zooms around the time of the burst and subsequent 
oscillations of the proto-neutron star. Correspondingly, the lower panel displays the results of the denoising using only 
the BBH dictionary with a value of $\lambda = 0.01$. Clearly each dictionary discriminates well between the type of 
signal it has been designed to search for, despite both signals overlap in time. When using the burst dictionary the method 
returns no BBH signal, as can be seen in the middle panel. Likewise, when using the BBH dictionary, no burst signal is 
visible in the lower panel, and the late inspiral and merger parts of the BBH signal are recovered properly. The discrepancies 
in the early inspiral and the spurious oscillations after the ringdown are to be expected, as we have explained before.

\subsection{Low SNR scenario}

The following test we consider is a low SNR scenario, namely SNR 6. Our strategy to denoise the signal in such a challenging situation consists in using the dictionary in combination with spectrograms, a common tool in data analysis. This test has two main goals. On the one hand, it allows us to check if our denoising procedure can improve the results of the spectrogram. On the other hand, we can test if the dictionary can recover the signal with acceptable accuracy in a low SNR scenario, once the time of arrival is known thanks to the spectrogram. We proceed as follows. First we apply the dictionary denoising with a generic value of $\lambda$, namely $\lambda=0.02$. Its value should be lower than that for SNR 20 to allow to recover the signal and also part of the noise. Then we calculate the spectrogram and select a window around the time of the maximum power (integrated over all frequencies). This step is a simple version of the event trigger generator implemented on the detectors \cite{Abbott:2016}. Next, we apply the iterative denoising procedure to this small window. In this case, we select the number of iterations that minimizes (maximizes) the MSE (SSIM) values. We have observed that the dependence of these values with the number of iterations does not follow a convex distribution. Therefore, it is unfortunately difficult to find a general rule which gives information on what is the best number of iterations. 

\begin{figure}[t]
  \centering
    \includegraphics[width=85mm]{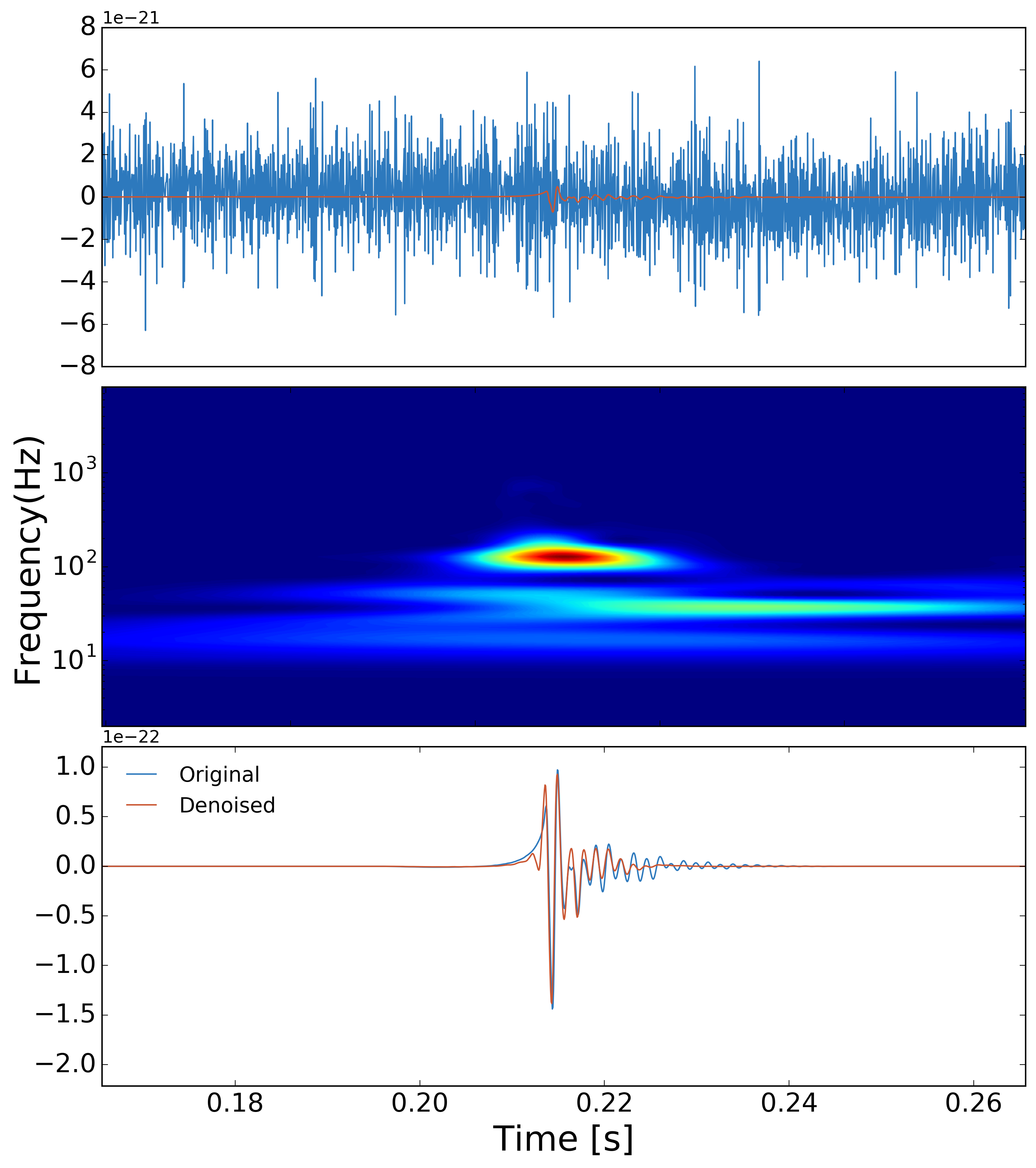}
  \caption{Denoising of signal \#1 of the burst catalog employing iterative denosing and spectrograms. The middle panel shows the spectrogram of the 0.5 s denoising with a general value of $\lambda$. The results shown at the lower panel were calculated using
  $\lambda=0.0095$ with $12$ iterations, and SNR 6.}
  \label{fig:low_snr}
\end{figure}

The results are shown in Fig.~\ref{fig:low_snr} for signal \#1 of the burst catalog. The middle panel of this figure displays the spectrogram. The denoised signal shown in the lower panel after applying our two-step procedure clearly indicates the
benefits of using the combined approach. The values of the MSE ($0.2\times10^{-3}$) and SSIM (0.88) indicate that the reconstruction is quite accurate even with this low SNR. These values are similar to those for SNR 10 reported in Table~\ref{Table:Burst}.  Therefore, the results of this test fulfill our two objectives. A key issue in this example is how to find the correct arrival time. We note that when using an even lower value of the SNR the arrival time obtained by the power integration does not always correspond to the arrival time of the signal. This issue could be solved by applying the iterative denoising directly to the list of candidate triggers.

\subsection{LASSO selection}

Up to now our experiments have focused on the goal of obtaining the best signal reconstruction for a generic value of the regularization parameter $\lambda$. In this section we show how the LASSO algorithm can also be used to infer some basic physical parameters of the sources from the denoised signals. We plan to further develop this approach towards parameter estimation in the near future.

We use the entire catalog as dictionary, except the test signals, without the learning procedure, and use the LASSO to select the signals closest to the one we use as test. The dictionary is not normalized in order to maintain the relative amplitude between the different waveforms that compose the catalog. Ideally, if a signal captured by the detector is inside the catalog, LASSO will select that signal. However, as the reconstruction is not perfect due to noise and the signal is not inside the catalog, LASSO will return the combination of signals that are more similar to the denoised one. 
To investigate if the selection can indeed be used to extract the physical parameters of the original signal, we devise the following procedure: firstly, we perform the denoising of the signal with a random time of arrival. We employ SNR 20 and $\lambda=0.03$. Secondly, once we have a clean signal, we use LASSO with the catalog and obtain the corresponding coefficients of each signal. It is possible to reconstruct the waveform using these coefficients and the catalog. Therefore, the value of $\lambda$ in the selection is the one that minimizes the error between the denoised signal and the reconstructed one.

As an example, the results of the parameter estimation of test signals \#1 and \#7 of the burst catalog are shown in Table~\ref{Table:parameters-burst}. The actual numbers of these two signals inside the catalog are \#26 and \#123 respectively.
In this table the signal listed above the horizontal line is the test signal, and the next three lines indicate the corresponding three signals of the catalog with the highest LASSO coefficients (employing the original numbering of signals of the catalog). 
This table shows that the physical parameters of the collapse progenitors are reasonably identified, especially for test
signal \#123 where the only discrepancy is on the EOS. (This is however to be expected as the catalog only uses two EOS, Shen and LS, and the signal for the LS EOS is the actual test signal and is hence {\it removed} from the catalog.) The estimations for signal \#26 are not as good but at least the differential rotation $A$ of the progenitor is well obtained. 

We turn next to estimate the parameters of test signal \#2 from the BBH catalog using the same procedure as for burst waveforms. Table~\ref{Table:parameters_BBH} reports the correspondence between the test signal and the three signals from the BBH catalog with the highest LASSO coefficients. Again, we find a good overall agreement, being the physical parameters of the test signal and the selected ones in a similar range. The main discrepancy is found in the BH masses, $M_1$ and $M_2$.

The mismatch observed in the parameter estimation is produced for two main reasons. On the one hand, 
the LASSO capabilities for parameter estimation obviously depend on how dense is the catalog. In the case of the 
two catalogs we employ in this work (and of most catalogs for that matter), the physical parameters are not sampled with detail due to the large computational cost of the simulations. It is likely that our results might improve when more complete catalogs become available. 
The second reason is due to the fact that we are using denoised waveforms as input, which includes the errors from the
dictionary reconstruction. The more accurate the deinoising results, the more precise the parameter estimation. However, since in a real world application the original signal is unknown, the LASSO classification can still be regarded as a useful tool to complement existing parameter estimation techniques  { (see e.g.~\cite{Jaranowski:2012,1606.01262,Logue:2012,cornish:2015})}.

\begin{table}
 \centering
  \caption{Parameter estimation: comparison between the physical parameters of test signals \#1 and \#7 of the burst catalog. From left to right the columns report: number of catalog signal, model name,  progenitor mass $M$, degree of differential rotation $A$,  precollapse angular velocity at the center $\Omega_{c,i}$,  precollapse rotation rate $\beta_i$, and equation of state. } \vspace{2mm}
  \begin{tabular}{@{}ccccccc@{}}
  \hline \hline
  Signal& Model name & $M$ &A&$\Omega_{c,i}$&$\beta_i$ &EOS\\
  &  &$[M_{\odot}]$&[$10^8$ cm]&[rad $\rm{s}^{-1}]$ &[\%]&\\
        \hline
        \hline
        \\	
        \#123 (\#7) & s40A3O12&40	& 0.5 & 10.65 & 1.84& LS \\      
\hline
     \#124 & s40A3O12&40	& 0.5 & 10.65 & 1.84& Shen \\    
     \#114 & s40A2O13&40	& 1.0 & 6.45 &2.60 & Shen\\       
     \#59 & s15A3O12	&15	& 0.5 & 10.65 & 1.60& LS \\   
     \\    
        \#26 (\#1) & s11A3O09&11	& 0.5 & 8.99 &0.72 & Shen \\       	
\hline
     \#24 & s11A3O07	&11 &0.5 & 5.95 & 0.40& Shen \\        
     \#54 & s15A3O05	&15	& 0.5 & 4.21 & 0.25& Shen \\       
     \#25 & s11A3O09	&11 &0.5 & 8.99 & 0.72& LS \\       
\label{Table:parameters-burst}
\end{tabular}

\end{table}

\begin{table}
 \centering
\caption{Parameter estimation: comparison between the physical parameters of BBH test signal \#2. From left to right the columns report: number of catalog signal, initial BH separation $r_0$ (in units of mass), initial orbital frequency $M\omega_0$, initial expansion factor 
$\dot{a}_0=\dot{r}_0/r_0$, Christodoulou masses of the two BHs at $t=0$, orbital eccentricity $\epsilon$, number of orbits between $t=0$ 
and common horizon time, and mass of final BH (remnant) $M$.}
 \vspace{2mm}
  \begin{tabular}{@{}ccccccccc@{}}
  \hline \hline
  Signal& $r_0$ & $M\omega_0$  &$\dot{a}_0$&$M_1$ &$M_2$ & $\epsilon$&Orbits& $M$ \\
        \hline \hline
        \\
             \#47 & 14 &0.017 &-0.00028 & 0.75 & 0.25 &  0.00047 & 22.7 & 0.96 \\    
	\hline
	     \#56 &15 & 0.015 & -0.00028 & 0.84 & 0.16 & 0.00049 & 28.8 & 0.98 \\      
	     \#29 &16 & 0.014 & -0.00033 & 0.60 & 0.40 & 0.00044 & 21.6 & 0.95 \\      
	     \#28 &16 & 0.014 & -0.00026 & 0.60 & 0.40 & 0.00016 & 23.8 & 0.94 \\      
\label{Table:parameters_BBH}
\end{tabular}

\end{table}
%

\section{GW150914}
\label{section:GW150914}

%
\begin{figure}[t]
  \centering
    \includegraphics[width=85mm]{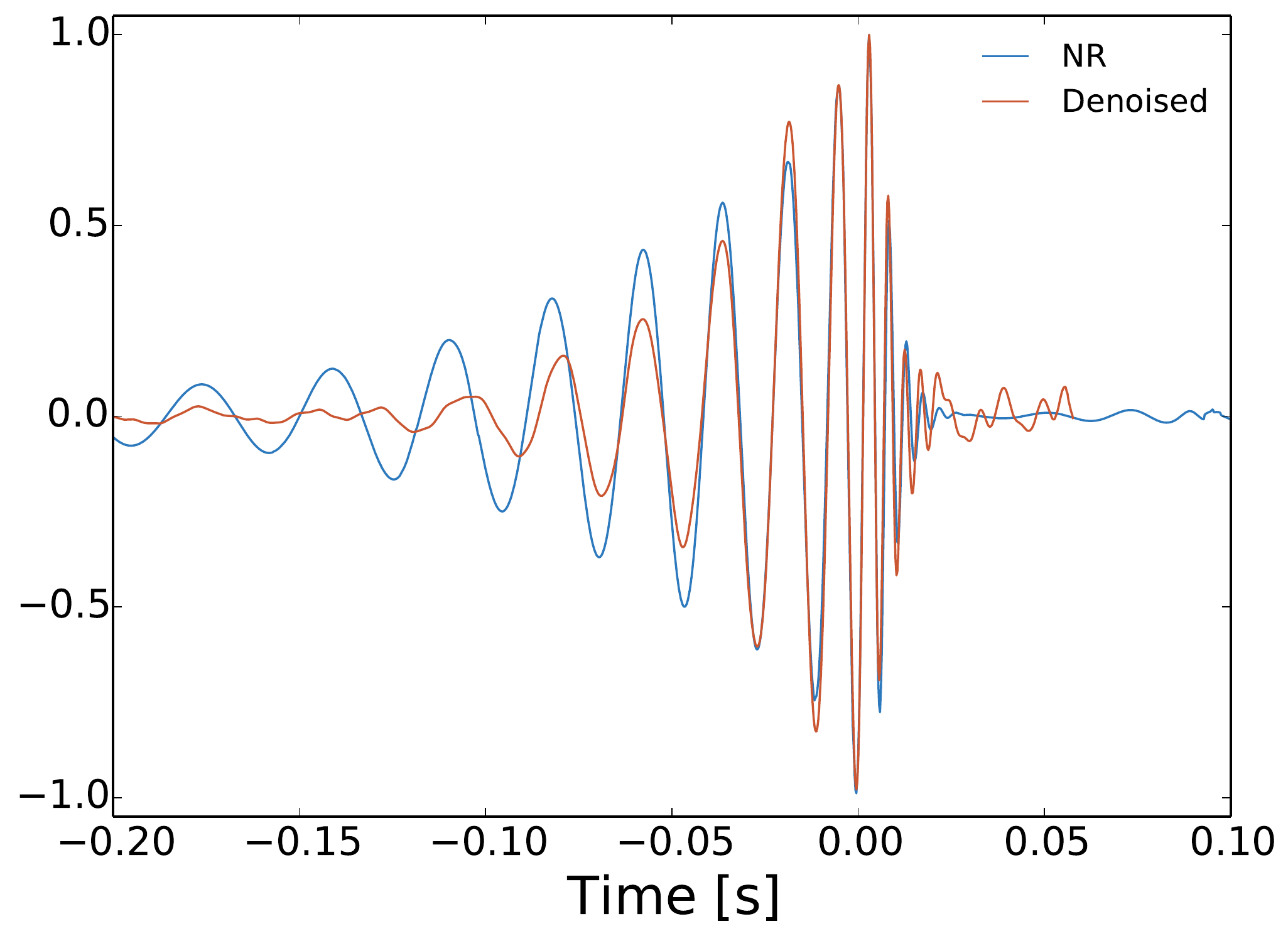}
  \caption{Denoising of signal GW150914 detected by advanced LIGO Hanford interferometer. We choose $\lambda=0.004$. The blue line indicates the NR template and the red curve corresponds to the actual signal. The amplitude of both signals has been rescaled to lie
  in the interval $[-1,1]$.}
  \label{fig:GW150914}
\end{figure}

For our final test we assess our algorithms with the real GW data of the discovery signal GW150914. We use as few assumptions about instrumental noise as possible, due to the fact that the detector noise is non-Gaussian and non-stationary. However, a minimum noise preprocessing is required due to two main reasons. On the one hand, there are well-known modeled sources of narrow-band noise (see Fig.~3 of~\cite{GW150914-prl}). On the other hand, ground-based detectors such as LIGO are not sensitive to low frequencies because of seismic noise. Therefore, we highpass the time series above 30 Hz to remove seismic noise and, following~\cite{GW150914-prl}, we also filter out all spectral lines. 

The results of applying the dictionary denoising procedure to the GW150914 Hanford signal are shown in Fig.~\ref{fig:GW150914}. 
The red curve in this figure displays the denoised signal obtained after applying the procedure to the real data. The implementation 
of the same approach to the best fit numerical relativity waveform~\cite{abbott-1} is shown with the blue curve. Even though the two 
signals are sampled at 4096 Hz, we use the high resolution dictionary as it leads to better results. This is common practice in the case 
of image denoising, where high resolution dictionaries are built to perform the denoising of low resolution images~\cite{Elad:2006}. 
Fig.~\ref{fig:GW150914} shows that the last cycles of the inspiral signal, the merger part, and the ringdown agree well with the NR 
waveform.  Comparing the two signals of Fig.~\ref{fig:GW150914} at $\pm0.15$ s from the minimum of the numerical relativity signal 
yields MSE = 0.0075 and SSIM = 0.4901. These quality measures show that while visually the comparison between both signals seems satisfactory, 
the reconstruction is not very accurate. A full parameter study  to find the optimal values of the dictionary algorithms for 
signals embedded in real noise will be presented elsewhere.  

\section{Summary and outlook}
\label{section:summary}

In this paper we have studied the capabilities of learned dictionaries to recover GW signals from a noise-dominated background. 
Our LASSO algorithm has been tested using signals from two main sources, bursts from rotational core collapse and chirps from BBH 
coalescence. To obtain the respective dictionaries, we have used 80\% of the waveforms for the training, 15\% for the validation, 
i.e.~to obtain the best set of parameters that produces the best results, and the last 5\% waveforms to assess the method.   
{ An interesting feature of LASSO is that, for most Gaussian noise realizations considered}, it returns zero if the input 
signal cannot be reconstructed by the atoms on the dictionary.  { As a result, the method may provide
a fairly clear signal reconstruction. On the other hand, an intrinsic limitation of the method is that the results} 
strongly depend on the selection of the regularization parameter $\lambda$, whose optimal value cannot be set a priori, and must 
be obtained with validation studies. It is possible that, if some noise transient (glitch) is similar to an actual GW signal, the method 
may be able to reproduce it and produce a false { reconstruction}. We defer for a future study the analysis of the false alarm rate 
using simulated (or real) glitches. We believe that this analysis, together with validation studies using real noise, are mandatory 
before using dictionaries in a detector's pipeline. To avoid false positives caused by glitches, it is possible to apply our method 
employing dictionaries built from a collection of known glitches. This idea follows the line of research proposed by~\cite{Powell:2015,Powell:2016}. 

Our results demonstrate that it is possible to extract GW signals embedded in Gaussian noise with good accuracy, using a generic 
dictionary with a common value of $\lambda$ for each type of waveforms. Overall, our results show that the denoising procedure 
works better for bursts than for BBH signals. This can be due to two main reasons. First, the morphology and the duration of both 
signals are very different. In particular, in the case of BBH signals, significantly longer than bursts, we need more atoms to cover 
all the signal duration. The second 
reason is related to the trained dictionary itself. The atoms on the burst dictionary are quite different between them which allows to 
cover more signal morphologies with a combination of only a few of them (sparse representation). In contrast, the atoms on the BBH 
dictionary are much more homogeneous since the signals used to train the dictionary are similar in the inspiral phase and only 
differ more clearly at the merger and the ringdown parts. Therefore, most of the atoms of the BBH dictionary cover the inspiral part, 
which can be reconstructed more easily. The larger inaccuracies appear when recovering the merger and the ringdown signals 
because there are less atoms to cover these parts and the reconstruction is less adaptable. A possible solution to this issue could 
be to use a couple of dictionaries, one to cover the inspiral part and a second one to cover only the merger and the ringdown parts. 
This may be worth investigating in the future. In addition, we have shown that using the LASSO algorithm iteratively can improve 
the results. Once a collection of triggers (i.e.~arrival times) is available, it is possible to obtain the signal in a few iterations even for
low SNR values. However, in order to determine the number of iterations that produces the optimal results, a more detailed study is 
necessary. 

We have also reported results on the use of the LASSO algorithm as a classification method (i.e.~for parameter estimation). The 
classification depends on how dense is the catalog and on how much noise can be removed from the original signal. The results 
become more accurate the larger the collection of waveforms available in the catalogs and the larger the physical parameters those 
catalogs cover, which is a major computational task. In particular, in the case of burst signals from core-collapse, the computational 
cost involved in calculating the GW waveforms renders unfeasible to obtain a large enough template bank. This classification method 
has therefore the same limitations than matched filtering. Even so, using the LASSO algorithm as a classification method deserves 
attention, particularly if used jointly with matched filtering techniques. Finally, we have also briefly shown the performance of 
dictionary-learning techniques for actual GW signals under real noise conditions. The results for the discovery signal GW150914 
seem promising.

There exist a large variety of learning techniques in the literature. In the present work we have only considered one specific
method but in the near future we plan to implement additional methods to perform the learning and to
compute the LASSO algorithm more efficiently. Obtaining the denoised solution for one patch of 256 samples takes typically a 
few tens of ms on an Apple iMac computer with Intel Core i7 processor and 16 Gb of Ram. The most expensive computational cost 
is associated with the learning task. Reducing the time involved in this part of the method is a key issue in order to eventually apply 
the method in real time to the actual data generated by the detectors. In the next few months, advanced LIGO and advanced Virgo will (re)start 
observing runs with improved sensitivity, increasing the number of detections. The development of sophisticated data analysis 
techniques to improve the opportunities of detection, especially for low SNR events, is therefore a most crucial effort.
{  The study reported in this work has shown that if the data are in good enough agreement  with the morphology of the 
atoms used to produce the dictionary, dictionary-learning algorithms may be used to extract signals from noise and to infer 
physical parameters. These algorithms could thus be a complementary addition to the gravitational wave data analysis toolkit.}

\section*{Acknowledgements}
{ We  thank  the  anonymous  referee  whose constructive review helped to improve
the manuscript.} Work supported by the Spanish MINECO (grants AYA2013-40979-P and 
MTM2014-56218-C2-2-P) and by the Generalitat Valenciana (PROMETEOII-2014-069).	

\appendix
\section{Correspondence with signal catalogs} \label{App:AppendixA}

For the sake of completeness and to facilitate the identification of the NR waveforms used in this study, Table~\ref{Table:parameters} 
reports the correspondence of the waveforms of our two dictionaries with the original naming of the burst~\cite{Dimmelmeier:2008} and 
BBH~\cite{Mroue:2013} signal catalogs.

\begin{table*}
 \centering
  \caption{Relation between the number of the GW signals employed in the validation set (first two columns) and in the test set 
  (last two columns) with the corresponding signal in the core-collapse catalog~\cite{Dimmelmeier:2008} and in the BBH 
  catalog~\cite{Mroue:2013}.} \vspace{2mm}
  \begin{tabular}{@{}cccccc@{}}
  \hline \hline
      & \multicolumn{2}{c}{Validation}&&\multicolumn{2}{c}{Test} \\
   	\cline{2-3} \cline{5-6}
 	\# & Core Collapse & BBH (SXS)&& Core Collapse & BBH (SXS) \\
        \hline
        \hline
        \#1&S15A2O13\_shen&SXS:BBH:0022 & &S11A3O09\_shen &SXS:BBH:0030\\
        \#2& S15A3O07\_ls&SXS:BBH:0003& &S15A1O09\_ls &SXS:BBH:0047\\  
        \#3&S15A3O15\_shen&SXS:BBH:0028 & &S15A1O09\_shen &SXS:BBH:0068\\  
        \#4& S15A2O13\_ls&SXS:BBH:0076& &S15A3O15\_ls &SXS:BBH:0087\\
        
        \#5&S15A1O01\_ls&SXS:BBH:0001 & &S20A2O07\_shen &\\ 
        \#6&S20A1O09\_shen&SXS:BBH:0077 & &S40A1O01\_shen &\\
        \#7&S11A1O01\_shen&SXS:BBH:0090 & &S40A3O05\_ls &\\
        \#8& S40A2O15\_ls&SXS:BBH:0053 & &S40A3O12\_ls &\\
        \#9& S20A2O05\_ls&SXS:BBH:0019& & &\\
        \#10& S20A1O09\_ls&SXS:BBH:0041 & & &\\
        
        \#11&S15A2O07\_ls&SXS:BBH:0091 & & &\\
        \#12& S20A3O07\_ls&SXS:BBH:0050 & & &\\
        \#13& S11A2O07\_ls&SXS:BBH:0080 & & &\\
        \#14&S11A1O13\_ls&SXS:BBH:0062 & & &\\   
        
        \#15&S15A2O05\_ls&SXS:BBH:0059 & & &\\
        \#16& S11A3O13\_shen& & & &\\   
        \#17& S11A3O09\_ls& & & &\\  
        \#18& S40A1O01\_ls& & & &\\ 
        \#19& S11A1O05\_shen& & & &\\
        \#20& S20A3O13\_shen& & & &\\ 

\label{Table:parameters}
\end{tabular}
\end{table*}



\end{document}